\documentclass[aps,pra,onecolumn,a4paper,10pt,showkeys,longbibliography]{revtex4-2}
\usepackage[english]{babel}
\usepackage[utf8x]{inputenc}
\usepackage[T1]{fontenc}

\usepackage[a4paper,top=3cm,bottom=2cm,margin = 25mm]{geometry}
\usepackage{bbold}

\usepackage{amsmath}
\usepackage{amssymb}
\usepackage{graphicx}
\usepackage{verbatim}
\usepackage{booktabs}
\usepackage{hyperref}
\usepackage{siunitx}
\usepackage{xcolor}

\newcommand{\ket}[1]{\left| #1 \right\rangle} %Makes a ket
\newcommand{\bra}[1]{\left\langle #1 \right|} %Makes a bra
\newcommand{\innProd}[2]{\left\langle #1 \middle| #2 \right\rangle} %Makes inner product
\newcommand{\diffD}{\mathrm{d}}

\newcommand{\matel}[2]{
\ifnum 0=#1\relax
S_{#2}
\else
S_{#2}^{*}
\fi
}
\newcommand{\vecB}[1]{\mathbf{#1}} %Boldface vector

\begin{document}
\title{Frustrated tunneling dynamics in ultrashort laser pulses}
\author{Edvin Olofsson, Stefanos Carlström and Jan Marcus Dahlström}
\email{marcus.dahlstrom@matfys.lth.se}
\affiliation{Department of Physics, Lund University, Box 118, SE-221 00 Lund, Sweden}

\begin{abstract}
    We study a model for frustrated tunneling ionization using ultrashort laser pulses. The model is based on the strong field approximation and it employs the saddle point approximation to predict quasiclassical trajectories that are captured on Rydberg states. We present a classification of the saddle-point solutions and explore their behavior as functions of angular momentum of the final state, as well as the carrier--envelope phase (CEP) of the laser pulse. We compare the final state population computed by the model to results obtained by numerical propagation of the time-dependent Schrödinger equation (TDSE) for the hydrogen atom. While we find qualitative agreement in the CEP dependence of the populations in principal quantum numbers, $n$, the populations to individual angular momentum channels, $\ell$, are found to be inconsistent between model and TDSE. Thus, our results show that improvements of the quasiclassical trajectories are in order for a quantitative model of frustrated tunneling ionizaiton. 
\end{abstract}

\maketitle

\section{Introduction}\label{sec:Introduction}
Frustrated tunneling (FT) is a strong-field phenomenon that limits the ionization of atoms in intense low-frequency laser fields due to a recapture process of photoelectrons on Rydberg states \cite{FT_original}. This is reminiscent of \emph{stabilization} which has previously been studied in various regimes \cite{Gavrila2002,Eichmann2009,Morales2011,Kjellsson2017b}; in the tunneling regime, FT is a mechanism for populating Rydberg states which are stable against further ionization. While the overall population to Rydberg states can be understood in terms of Monte Carlo simulations of classical electrons initiated by tunneling ionization \cite{FT_original,Shvetsov_2009}, a true quantum description of the excitation process may reveal additional phenomena, e.g. quantum interference and diffusion, as well as competing multi-photon resonant paths \cite{Muller_1992,Zimmerman_2017}. Numerical simulations of strong-field excitation have been performed by propagating the time-dependent Schrödinger equation (TDSE) to evidence a strong dependence on both the laser pulse duration and the carrier--envelope phase (CEP), but the role of FT in ultrashort laser pulses has remained under debate \cite{Chen_2012}. 

Related processes, such as strong-field ionization, high-order harmonic generation (HHG) \cite{HHG_Lewenstein}, and high-order above-threshold ionization \cite{Becker_2002} have been described using the strong field approximation (SFA), which has helped to interpret experimental and numerical results in terms of Feynman-like path integrals.   
Within this framework, the saddle-point approximation can be employed to deal with oscillatory integrals that arise with complex saddle points as a result. Saddle points for the temporal integrals can be interpreted as initial and return times for quasi-classical electron trajectories, with the imaginary part reminding us that quantum effects, such as tunneling, are at play. Thus, the path-integral formulation provides a useful stepping stone between the classical and quantum description of the processes.  
Motivated by the success of such approaches, there have been recent attempts to explain FT within the SFA framework~\cite{Popruzhenko_FT,Hu:19}. An important issue arises as to how the excited states can be incorporated into the model. This is because the SFA only includes the initial state and the continuum states in the form of Volkov waves, and it does not include any bound excited states. 

The first goal of our investigation is to review the present SFA-FT models in a common framework of Dyson equations, our second goal is to classify the SFA-FT solutions and our third goal is to investigate how the CEP of ultrashort laser pulses affects the excited state population.  In order to support our model study, we present FT results from TDSE that are resolved in both $n$ and $\ell$, because this type of data has been missing in the literature and it has been proposed as a key to better understand interference effects in FT \cite{Popruzhenko_FT}. CEP modulations for a continuous range of CEPs in the population of excited states have been considered previously  in the multiphoton regime \cite{Nakajima_2006_1,Nakajima_2006_2}. Here, we consider the tunneling regime, where only results for a small set of CEP values have been presented, see e.g.\ Ref.~\cite{Chen_2012}. In addition to studying CEP modulations, we also investigate the role of the continuum states in populating excited states with TDSE.

The article is structured as follows: In Sec.~\ref{sec:Theory} an overview of present SFA-FT theories is presented. A common Dyson formalism in presented in Sec.~\ref{sec:dyson}, followed by a discussion on the role of quantum diffusion in Sec.~\ref{sec:quantumdiffusion} and a derivation of the complex SFA-FT interaction time in Sec.~\ref{sec:interactiontime}. In Sec.~\ref{sec:results} numerical saddle-point solutions for SFA-FT in ultrashort pulses are presented, with a detailed classification of solutions in Sec.~\ref{sec:Solutions} and a comparison of Rydberg state populations from TDSE in Sec.~\ref{sec:Populations}. Conclusion and outlook is given in Sec.~\ref{sec:conclusion}.

\section{Theory}\label{sec:Theory}
The strong-field excitation amplitude of an atom by a time-dependent laser pulse can be formally defined as: 
\begin{equation}\label{eq:Rydberg_amp}
   a_{n\ell m} =\lim_{t\rightarrow \infty}  \innProd{\psi_{n\ell m}(t)}{\Psi(t)},
\end{equation}
where $\ket{\psi_{n\ell m}(t)}=\ket{\phi_{n\ell m}}\exp(-i\varepsilon_{n\ell m}t)$ is the targeted Rydberg state of the field-free atom and the electron wave packet, 
\begin{equation}
    \ket{\Psi(t)} = U(t,t_i)\ket{\psi_0(t_i)} \rightarrow \sum_{n\ell m} a_{n\ell m}\ket{\psi_{n\ell m}(t)}+\int d^3k \, a_{\vecB{k}}\ket{\psi_{\vecB{k}}(t)}, \, t\rightarrow\infty
\end{equation}
represents the time evolution from the initial field-free atomic state, $\ket{\psi_0(t_i)}=\ket{\phi_0}\exp[iI_p t_i]$ (quantum numbers $n_0=1$, $\ell_0=0$ and $m_0=0$ for the hydrogen atom) with $U(t,t_i)$, being the full propagator describing the combined action of the field-free atomic Hamiltonian, $H_A$, and the light--matter interaction, $V_L(t)$.
The excitation amplitudes converge to constant values to all eigenstates of $H_A$, including both the bound part: $\ket{\phi_{n\ell m}}$ and continuum part: $\ket{\phi_{\vecB{k}}}$.  
In this work we define the vector potential of the laser pulse as 
\begin{equation}
    A(t) = A_0\cos^2\left(\frac{\pi t}{\tau}\right)\sin(\omega t + \varphi),
    \label{eq:Alaserpulse}
\end{equation}
with an electric field defined as $E(t)=-\partial A/\partial t$.  
The fields are restricted to $-\tau/2<t< \tau/2$, such that $\tau$ is the entire duration  of the laser pulse.  
All calculations are performed within the dipole approximation with
the light-matter interaction expressed in length gauge, $V_L^\mathrm{(Len.)}(t)=zE(t)$, or velocity gauge, $V_L^\mathrm{(Vel.)}(t)=p_zA(t)+A^2(t)/2$. Atomic units are used unless otherwise stated and the atomic ionization potential is denoted by $I_p > 0$. 

\subsection{Dyson equations for frustrated tunneling}\label{sec:dyson}
In this subsection we review the SFA-FT theories presented in Refs.\ \cite{Popruzhenko_FT,Hu:19} in the language of Dyson equations. Consider a Hamiltonian $H(t)$ that can be separated in two parts, $H_0(t)$ and $V(t)$, where the exact form of the time evolution operator associated to $H_0$ is known, $U_0(t,t_i)$. Then one can write down a recursive solution for the time evolution operator associated to the full Hamiltonian as: 
\begin{equation}\label{eq:Dyson}
    U(t,t_i) = U_0(t,t_i)-i\int_{t_i}^t\diffD t' U(t,t')V(t')U_0(t',t_i).
\end{equation}
Using this Dyson equation for $U$, one can rewrite the exact amplitude for strong-field excitation, Eq. \eqref{eq:Rydberg_amp}, as: 
 \begin{equation}\label{eq:Rydberg_amp_Pop}
    a_{n\ell m} = \lim_{t\rightarrow \infty}-i\int_{t_i}^t\diffD t' \bra{\psi_{n\ell m}(t)}U(t,t') V_L(t')U_A(t',t_i)\ket{\psi_0(t_i)},    
 \end{equation}
 with $H_0=H_A$ being the atomic Hamiltonian, and $V(t) = V_L(t)$ being the light--matter interaction. 
Alternatively, the exact amplitude, Eq.~(\ref{eq:Rydberg_amp_Pop}), can be rewritten by an additional application of the Dyson equation, with the atomic potential, $V(t)=V_A$, as a second interaction to yield,
 \begin{align}\label{eq:Hu}
 \begin{split}
      a_{n\ell m} &= \lim_{t\rightarrow \infty}\bigg[-i \int_{t_i}^{t}\diffD t' \bra{\psi_{n\ell m}(t)}U_L(t,t')V_L(t')U_A(t',t_i)\ket{\psi_0(t_i)}\\ &
     +(-i)^2\int_{t_i}^{t}\diffD t''\int_{t_i}^{t''}  \diffD t'\bra{\psi_{n\ell m}(t)}U(t,t'')V_AU_L(t'',t')V_L(t')U_A(t',t_i)\ket{\psi_0(t_i)}\bigg].
   \end{split}  
 \end{align} 
 It should be noted that even if the recursive Dyson procedure is continued in this way, $U$ will always appear in the last term of the expansion. Assuming that the exact form of $U$ is not known, this implies that some approximation of the exact propagator must always be made in analytical calculations. 
 In Ref. \cite{Hu:19}, Hu et al. use Eq. \eqref{eq:Hu} and approximate $U$ by a propagator for dressed Rydberg states, $U_{n\ell m}$. Formally, the FT process is then a second-order process that requires two interactions: (i) an excitation by the laser field at $t'$ and (ii) a capture by the atomic potential at $t''>t'$. This is because the first-order term with $U_L$ vanishes in Eq.~(\ref{eq:Hu}), due to quantum diffusion in the limit $t\rightarrow\infty$ \cite{Hu:19}. In this picture, the capture can happen at any time after the laser excitation, $t''>t'$, while in reality the population to Rydberg states via FT has to occur at a finite time before the end of the laser pulse.   
 Less formally, first-order theories for FT can be developed by applying different approximations to $U$ in Eq.~(\ref{eq:Rydberg_amp_Pop}) for different times based on physical intuition. For instance, one could approximate $U$ by $U_L$ when the field is strong, and then switch back to $U_A$ when the field is weak. Provided that the laser field ends abruptly at finite time, $t=t_f$, Popruzhenko has proposed that the SFA momenta can be directly matched to Rydberg states using the Laplace–Runge–Lenz vector (LRL) \cite{Popruzhenko_FT}. In general, the ``end'' of the laser pulse is not always possible to define uniquely, but for the ultrashort laser pulses of $\cos^2$-type that are considered in this work, defined  Eq.~(\ref{eq:Alaserpulse}), the field ends at $t_f=\tau/2$. This makes it possible to test the Popruzhenko SFA-FT theory with physical pulses and compare to Rydberg populations from exact TDSE simulations in hydrogen.

 In order to relate the exact Dyson solution with the SFA theory at $t_f$, the exact amplitude for strong-field excitation in Eq.~(\ref{eq:Rydberg_amp}) can be written in terms of plane-wave momentum states using the identity,  
${\mathbb{1}}=\int \diffD^3p \ket{\vecB{p}}\bra{\vecB{p}}$, as
\begin{align}
\begin{split}
   a_{n\ell m}(t_f) &=  \int\diffD^3p \innProd{\psi_{n\ell m}(t_f)}{\vecB{p}}a(\vecB{p},t_f), 
\label{eq:anlmp}
 \end{split}
\end{align}
where
\begin{align}
a(\vecB{p},t_f)&=-i\int_{t_i}^{t_f}\diffD t' \bra{\vecB{p}}U(t_f,t') V_L(t')U_A(t',t_i)\ket{\psi_0(t_i)}. 
\label{eq:aptf}
\end{align}
While Eq.~(\ref{eq:anlmp}) is exact and will approach a constant value for $a_{n\ell m}(t_f\rightarrow\infty)$, it is worth to note that this formulation is not ``well behaved'' due to $\ket{\vecB{p}}$ not being an eigenbasis of $H_A$. 
As a result, the intermediate projection amplitude, $a(\vecB{p},t_f)$, does not approach a constant value as $t_f\rightarrow\infty$.   
 
 In the following, we have chosen to study the SFA-FT theory of Ref.~\cite{Popruzhenko_FT} due its two main advantages:  
 (i) the model is based on a first-order interaction with the laser field and 
 (ii) all momentum variables are eliminated by substitution using the LRL vector. 

\subsection{Quantum diffusion}\label{sec:quantumdiffusion}
From Eqs.~(\ref{eq:anlmp}) and (\ref{eq:aptf}) it is evident that integration over both plane-wave momenta, $\vecB{p}$, and interaction times, $t'$, must be performed in order to establish the SFA-FT amplitude. In saddle-point approaches, the order of integration is not important \cite{Smirnova_Ivanov_book}, and we first consider the momentum integration separately. Integration of momentum was omitted in prior work \cite{Popruzhenko_FT}. 
To proceed, we replace $a(\mathbf{p},t_f)$ in Eq.~(\ref{eq:anlmp}) by the SFA amplitude for photoionization 
\begin{equation}\label{eq:SFA_amplitude}
a_\mathrm{SFA}(\vecB{p},t_f) = 
 -i\int_{t_i}^{t_f}\diffD t' \bra{\vecB{p}}U_L(t_f,t') V_L(t')U_A(t',t_i)\ket{\psi_0(t_i)}  
\end{equation}
and perform a saddle point integration over the momentum space, 
\begin{align}
\begin{split}
   a_{n\ell m}(t_f) & 
   \approx \sum_{\vecB{p}_s}
   \innProd{\psi_{n\ell m}(t_f)}{\vecB{p}_s}a_\mathrm{SFA}(\vecB{p}_s,t_f), 
\label{eq:anlmSFA}
 \end{split}
\end{align}
where the stationary momentum contributions are
\begin{align}
\begin{split}
   a_{\mathrm{SFA}}(\vecB{p}_s,t_f) & 
   \approx -i\int_{t_i}^{t_f}\diffD t'
   \left(
   \frac{2\pi i}{t_f-t'}
   \right)^{3/2}
   \bra{\vecB{p}_s}U_L(t_f,t') V_L(t')U_A(t',t_i)\ket{\psi_0(t_i)}.
\label{eq:aSFAps}
 \end{split}
\end{align}
In writing Eqs.~(\ref{eq:anlmSFA}) and (\ref{eq:aSFAps}), we  have used the SFA action, 
\begin{equation}\label{eq:Action}
S_0(t',\vecB{p}) = \int_{t'}^{t_f}\diffD t'' \frac{1}{2}[\vecB{p} + \vecB{A}(t'')]^2 - I_p(t'-t_i), 
\end{equation}
and we have assumed that the momentum dependence on $\vecB{p}$ of the Rydberg state,  
$\innProd{\phi_{n\ell m}}{\vecB{p}}$, is small as compared to the SFA amplitude, 
$a_\mathrm{SFA}(\vecB{p},t_f)$. Rough estimations can be made using the width of the Rydberg state: $\Delta p^\mathrm{Ryd} \approx 2/\Delta r^\mathrm{Ryd} \approx n^{-2}$, while the SFA width is: $\Delta p^{\mathrm{SFA}}\approx(t_f-t')^{-1/2}$. Physically, the window of capture by the Rydberg state in space should be small compared to the size of the returning SFA wave packet.  
The momentum projection, $\innProd{\phi_{n\ell m}(t_f)}{\vecB{p}_s}$, can be approximated by its angular factor, $ \innProd{\phi_{n\ell m}(t_f)}{\vecB{p}_s}\approx Y_{\ell m}^*({\vecB{p}/\sqrt{{p}^2}})$ or evaluated using its analytical form, which is known for hydrogen \cite{Bethe1977}. A direct comparison of these prefactors will be shown in Sec.~\ref{sec:Populations}. 
Clearly, Eq.~(\ref{eq:anlmSFA}) can be interpreted as FT because the electron goes from the ground state to a Rydberg state via the continuum states. The laser-driven dynamics is associated with a quantum diffusion factor, $(t_f-t')^{-3/2}$, that strongly damps long trajectories in the continuum.   
In SFA the condition of stationary momentum, $\vecB{p}=\vecB{p}_s(t_f,t')$, is that the electron returns to the atom, 
\begin{align}
\vecB{r}_s(t_f)=\int_{t'}^{t_f} \diffD \tau [\vecB{p}_s+\vecB{A}(\tau)] = (t_f-t')\vecB{p}_s+\Delta G(t_f,t') = 0,  
    \label{eq:pSFAreturn}
\end{align}
which is a reasonable condition for a recapture process by the atom.  
Improvements to the recapture process on complex targets can be studied by adding a constant term, $\vecB{\Delta r}_{n\ell }$, on the right hand side of Eq.~(\ref{eq:pSFAreturn}). This implies a shift of the stationary momentum,  
\begin{align}
\vecB{p}_{s}=-\Delta G(t_f,t') \rightarrow \vecB{p}_{n\ell ,s}=
(-\Delta G(t_f,t')+\vecB{\Delta r}_{n\ell })/(t_f-t'), 
\end{align} 
such that the final position, $\vecB{r}_{n\ell ,s}(t_f)=\vecB{\Delta r}_{n\ell }$, is reached within any given time window, $t_f-t'$. 
This ad-hoc modification does not affect the quantum diffusion factor in Eq.~(\ref{eq:aSFAps}), and it can be motivated by exterior centres of Rydberg states (for $\ell>0$) \cite{Hu:19}. In the next subsection, we will show that the momentum substitution by the LRL-vector, proposed by Popruzhenko \cite{Popruzhenko_FT}, leads to quasi-classical trajectories that do not return exactly to the atomic core at the end of the laser pulse.

\subsection{Complex interaction time}\label{sec:interactiontime}
In SFA the integral over interaction times, $t'$ in Eq.~(\ref{eq:SFA_amplitude}), is evaluated using the saddle point method for which the action at time $t'=t_s(\vecB{p})$ is stationary by the tunneling condition, 
\begin{equation}
\label{eq:Saddle}
\left[ \vecB{p} + \vecB{A}(t_s) \right]^2 = -2I_p.
\end{equation}
The saddle point contributions are then weighted by the factor, 
\begin{equation}
[\diffD^2S/\diffD t'^2(t_s)]^{-\eta/2-1/2}=[\vecB{p}+\vecB{A}(t_s)]^{-\eta/2-1/2},
\end{equation}
where $\eta=Z/\sqrt{2I_p}$ is an effective principal quantum number of the initial state, c.f. Ref.~\cite{Popruzhenko_FT}. 
The electron then follows a stationary trajectory specified by
\begin{equation}\label{eq:Traj}
\vecB{r}_s(\vecB{p},t)  
= \vecB{p}(t-t_s) + \Delta \vecB{G}(t,t_s),
\end{equation}
which has the following properties:
\begin{equation}
\vecB{r}_s(\vecB{p},t_s) = 0,  
\quad
\dot{\vecB{r}}_s^2(\vecB{p},t_s) = -2I_p,
\quad
\dot{\vecB{r}}_s(\vecB{p},t_f) = \vecB{p}.
\label{eq:trajcond}
\end{equation}
where the first and second identities in Eq.~(\ref{eq:trajcond}) state that the electron starts at the origin with imaginary kinetic momentum at the complex interaction time, $t_s$, while the third identity states that the final kinetic and canonical momenta are equal at the end of the pulse. 

Following Popruzhenko \cite{Popruzhenko_FT}, we now apply a constraint on the momentum, $\vecB{p}\rightarrow\vecB{p}_{n\ell ,s}$, by imposing that the final squared angular momentum, 
\begin{equation}\label{eq:AngMom}
L^2(t_f) = [\vecB{r}_s(\vecB{p},t_f)\times\vecB{p}]^2 = p_{\perp}^2\Delta G(t_f,t_s)^2 = \ell(\ell+1),
\end{equation}
of the stationary trajectory has the quantized squared angular momentum, $\ell(\ell+1)$, of the target Rydberg state. 
Given that the angular momentum condition, Eq.~(\ref{eq:AngMom}), determines the square of the transverse momentum component, 
\begin{equation}
p_{\perp}^2 = \frac{\ell(\ell+1)}{\Delta G(t_f,t_s)^2},
\end{equation}
the component $p_z$ parallel to the laser field can now be solved using Eq.\ \eqref{eq:Saddle} as  
\begin{equation}
p_z = -A(t_s) \pm i \sqrt{2I_p + p_{\perp}^2} = -A(t_s) \pm i \sqrt{2I_p + \frac{\ell(\ell+1)}{\Delta G(t_f,t_s)^2}}.
\end{equation}
In this way the stationary momentum has been eliminated and one remaining condition is required to solve for $t_s$. 
Based on the LRL-vector, a condition on the final energy is imposed,
\begin{equation}\label{eq:EnCons}
\frac{\vecB{p}^2}{2} -\frac{Z}{\sqrt{\vecB{r}_s(\vecB{p},t_f)^2}} = -\frac{Z^2}{2n^2}, 
\end{equation}
that introduces a dependence on the principal quantum number, $n$, of the Rydberg state. 
The Rydberg matching conditions, Eqs.~(\ref{eq:AngMom}) and (\ref{eq:EnCons}), are used instead to the condition for stationary momentum, Eq.~(\ref{eq:pSFAreturn}).  

In order to find approximate solutions to the SFA-FT model, 
we reformulate Eq.~(\ref{eq:EnCons}) as 
\begin{align}
    \label{eq:EnConReform}
    A(t_s)=\pm \sqrt{\frac{2Z}{\sqrt{\vecB{r}_s^2}}-\frac{Z^2}{n^2}-\frac{\ell(\ell+1)}{\Delta  G_s^2}}\pm i \sqrt{2I_p+\frac{\ell(\ell+1)}{\Delta G^2_s}},
\end{align}
where $\vecB{p}$ has been eliminated and $\vecB{r}_s$ and $\Delta G_s$ are functions of $t_s$. %
Without loss of generality, we consider an electric field peak at $t=0$, $E(t)\approx -E_0\cos\omega t$. 
Solutions to Eq.~(\ref{eq:EnConReform}) are then expected to occur for $t_s\approx 0$, such that $A(t_s)\approx A_0\sin(\omega t_s)\approx \omega A_0 t_s$. 
If the right hand side of Eq.~(\ref{eq:EnConReform}) is approximated with the zeroth-order interaction time, $t_s^0=0$, the first-order solutions become   
\begin{align}
    \label{eq:EnConReformApprox}
    t_s=
    \frac{1}{\omega A_0}\left(
    \pm 
    \sqrt{\frac{2Z}{r_0}-\frac{Z^2}{n^2}-\frac{\ell(\ell+1)}{\Delta  G_0^2}}
    \pm 
    i \sqrt{2I_p+\frac{\ell(\ell+1)}{\Delta G^2_0}}
    \right).
\end{align}
Here, the second term  is always imaginary and it describes the tunneling process (the positive sign must be chosen for damping). In contrast, the first term describes the recapture process, and it may be either real or imaginary depending on the final radial position of the electron, $r_0=\sqrt{\vecB{r}^2_0}$. 
Real contributions to $t_s$ are found for 
\begin{equation}
\label{eq:ReturnToRydberg}
r_0<\frac{2n^2}{Z + n^2 \ell(\ell+1)/(Z\Delta G_0^2)}<\frac{2n^2}{Z},
\end{equation}
which implies that the final electron position should not exceed a Kepler orbit with zero angular momentum. It is known from classical mechanics that increasing the angular momentum reduces the eccentricity of the orbit, and in the limit of large $n$ and $\ell$ the orbit becomes circular with radius $n^2/Z$. In the SFA-FT model we find that increasing the angular momentum, $\ell$, indeed decreases the critical radius for recapture.  This shows that the strict return condition for stationary momentum, given in Eq.~(\ref{eq:pSFAreturn}), has been relaxed, by Eqs.~(\ref{eq:AngMom}) and (\ref{eq:EnCons}), to include returns to the entire volume of the final Rydberg state. Provided that this return condition is satisfied, there are two solutions for $t_s$, one on either side of the electric field maximum, 
\begin{equation}
\label{eq:Rets}
\mathrm{Re}(t_s)=\pm\frac{\sqrt{2Z/r_0-Z^2/n^2-\ell(\ell+1)/\Delta G_0^2}}{\omega A_0}, 
\end{equation}
where the solution after (before) the electric field peak will be labeled $\alpha$ ($\beta$) in Sec.~\ref{sec:results}. 
Further, it can be seen that an increased angular momentum, $\ell$, decreases the real time delay $\Delta t_s$ between the $\alpha$ and $\beta$ solutions. 
At some critical angular momentum, $\ell>\ell_n$, the real contribution to $t_s$ is prevented, leading to an imaginary contribution also by the first term on the right hand side of Eq.~(\ref{eq:EnConReform}). This signals that the electron trajectory ends up outside the recapture volume of the Rydberg state. In this case the positive sign must be selected also for the first term in Eq.~(\ref{eq:EnConReformApprox}), such that the stationary time leads to suppression of the recombination probability for that trajectory. The critical angular momentum can be estimated as
\begin{equation}
    \label{eq:ln}
    \ell_n=-\frac{1}{2}+\sqrt{\frac{1}{4}+\Delta G^2_0\left(
    \frac{2Z}{\Delta G_0}-\frac{Z^2}{n^2}
    \right)},
\end{equation}
by setting the expression in the first square root of Eq.~\eqref{eq:EnConReform} to zero, and assuming that the motion is mostly due to the electric field so that $r_s$ can be approximated by $\Delta G_0=\Delta G(t_f,0)$. We note that this estimate gives a slightly smaller value for $\ell_{\infty}\approx 6.84$ compared to $\ell_{\infty}\approx 7.31$  as computed by Eq.~(33) in Ref.~\cite{Popruzhenko_FT} (for the pulse parameters considered in this article). 

In general, there will be multiple electric field peaks in ultrashort laser pulses, which lead to multiple sets of complex interaction times, $t_s$ that depend on the exact laser field. These solutions will be studied by solving Eq.~(\ref{eq:EnConReform}) numerically, including the self-consistent dependence of $t_s$ on the right hand side, in the next section.   

\section{Results}\label{sec:results}
In this section results for the SFA-FT model are presented with a direct comparison to exact results from the TDSE for the hydrogen atom. The model results complement and expand upon previous SFA-FT results \cite{Popruzhenko_FT}, because finite pulse duration, Rydberg state-projection factors and quantum diffusion effects are taken into account. This allows us to study FT, to Rydberg states with $n$ and $\ell$ quantum numbers, as a function of CEP for ultrashort laser pulses. We show that the SFA-FT trajectories can be classified by a complex filter and related to quasi-classical trajectories. Transitions between classically allowed and forbidden domains for recapture are found as the CEP is varied. Finally, TDSE simulations are presented where the continuum state propagation is damped in order to verify that FT is indeed the correct physical interpretation for $n\ge4$ with the considered laser parameters.  

\subsection{Solutions to the Popruzhenko FT theory for ultrashort pulses}
\label{sec:Solutions}
In this subsection we show solutions $t_s$ to Eq.~(\ref{eq:EnConReform}) that are obtained using Newton's method. ultrashort laser pulses, with vector potential given in Eq.~(\ref{eq:Alaserpulse}) for $A_0=1.147$~a.u., $\tau=2T$ and period $T=2.67$~fs, are considered. This corresponds to a full width at half maximum (FWHM) duration of $1.94$~fs and $1.5\times10^{14}$~W/cm$^2$ peak intensity.  

\subsubsection{Dependence on angular momentum of Rydberg state}
\begin{figure}[ht!]
\centering
    \includegraphics[scale = .375]{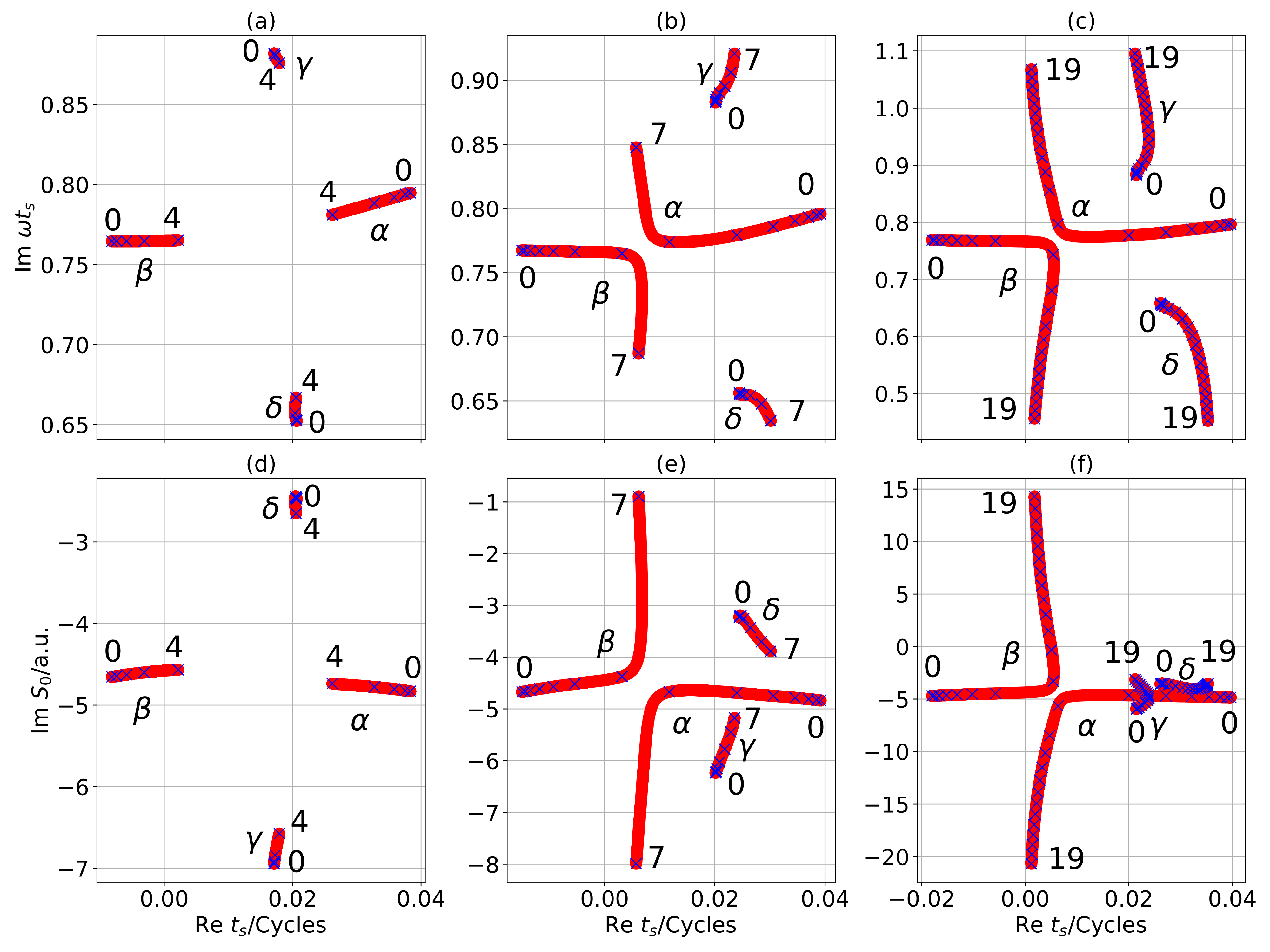}
    \caption{$\ell$ dependence of $t_s$ and $\textrm{Im}\{S_0\}$ for $n=5,8,20$. The numbers in each figure mark the start and end values of $\ell$. The blue crosses mark integer values of $\ell$. For $n=8,20$ the avoided collisions occur for $\ell$  between 6 and 7;  cf.\ discussion around Eq.~\eqref{eq:ln}.}\label{fig:l_dep_cont}
\end{figure}

\begin{figure}[ht!]
\centering
    \includegraphics[scale = .4]{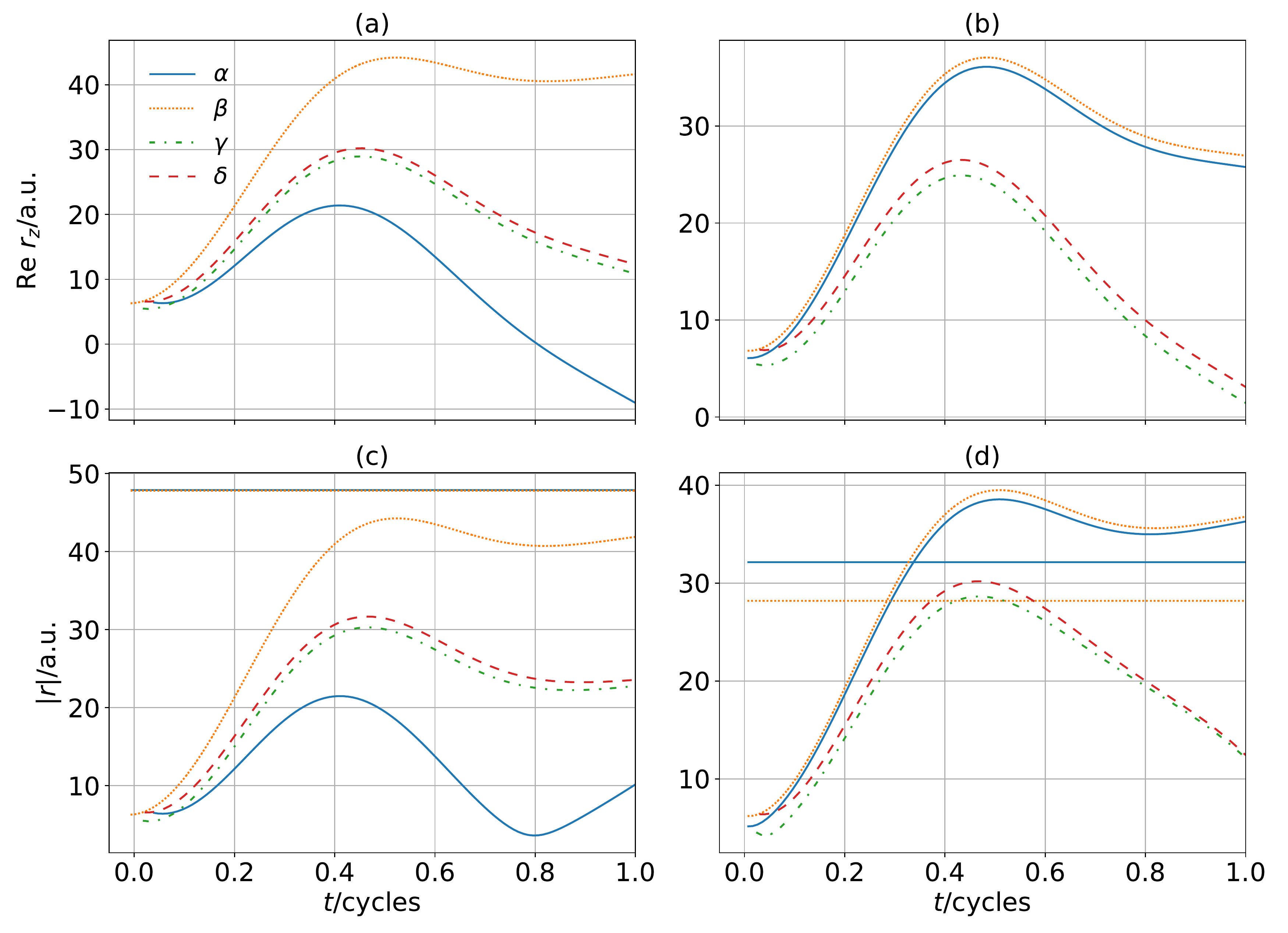}
    \caption{The top row displays the real part of the $z$-component of the trajectories associated to the four classes of solutions discussed in the main text, and the bottom row shows the absolute value of the (complex) position vector, for the same trajectories. The horizontal lines indicate the absolute values of the middle expression in Eq.~\eqref{eq:ReturnToRydberg} for the $\alpha$ and $\beta$ classes of solutions. (a) and (c) use $(n,l) = (5,1)$, while (b) and (d) uses $(n,l) = (8,7)$.}\label{fig:trajectories}
\end{figure}

In Fig.~\ref{fig:l_dep_cont} we show the $\ell$ dependence of the stationary interaction time, $t_s$, for a hydrogen atom to a Rydberg state with principal quantum numbers $n=5$ (a) $n=8$ (b) and $n=20$ (c), that is subject to an ultrashort laser pulse with cosine-like electric field ($\varphi=0$). The curves for $t_s$ are computed by continuous increase of $\ell$ from $0$ to $n-1$ with the quantized values, $\ell=0,1,...,n-1$, marked by $+$ symbols.  
In Fig.~\ref{fig:l_dep_cont}~(a,b,c) we observe four classes of solutions for $t_s$, labeled by $\alpha$, $\beta$, $\gamma$ and $\delta$. The $\alpha$ and $\beta$ solutions start far apart on the real axis with similar magnitude on the imaginary axis for $\ell=0$. The two solutions become closer together on the real axis as $\ell$ is increased. Clearly the $\alpha$ and $\beta$ solutions correspond to the approximate solutions, $t_s$, for Eq.~(\ref{eq:EnConReform}) found earlier in Sec.~\ref{sec:interactiontime}.  Since $\alpha$ starts after $\beta$, we may refer to the two classes of solutions as the {\it short} and {\it long} FT trajectories, respectively. It should be noted that these trajectories are not the same as the short and long HHG trajectories that both start after the electric field peak and both return to the origin for recombination with the atom. Because the short FT trajectory ($\alpha$) is created after the electric field peak it is such a returning trajectory that may scatter with the atomic potential. The long FT trajectory ($\beta$) starts before the electric field peak, which implies that it has a net drift momentum away from the atom and never returns to the origin. Thus, the $\beta$ solutions are most likely well described by the SFA model, while the $\alpha$ solutions may be strongly distorted due to rescattering with the ion. This point is illustrated in Fig.~\ref{fig:trajectories}, where we show the trajectories associated to the different types of solutions displayed in Fig.~\ref{fig:l_dep_cont}~(a) and (b), for $n=5$, $\ell=1$ and $n=8$, $\ell = 7$. The trajectories are plotted from the exit point of tunneling, $\textrm{Re}\{t_s\}$, to the end of the pulse, $t_f$. In accordance with the discussion above, we see that in Fig.~\ref{fig:trajectories}~(a) the  $\alpha$ trajectory crosses $\textrm{Re}\{r_z\} = 0$, while the $\beta$ remains relatively far away from the core, but still within the recapture radius of $\sim$50 Bohr radii, estimated by  Eq.~(\ref{eq:ReturnToRydberg}) (the horizontal line in Fig.~\ref{fig:trajectories}~(c) and (d) indicate the tighter of the two bounds). In Fig.~\ref{fig:trajectories}~(b) and (d) on the other hand, we see that both $\alpha$ and $\beta$ end up outside the recapture volume.

It can be seen in    Fig.~\ref{fig:l_dep_cont}~(a) that the $\alpha$ and $\beta$ solutions do not ``collide'' in the complex plane for $n=5$ as $\ell$ is increased from $0$ to $4$. 
This is because the angular momentum is restricted to low values below the critical angular momentum, 
$\ell_5\approx 4.51 > 4$, 
estimated using Eq.~(\ref{eq:ln}) with $\Delta G_0=26.8$ Bohr radii.  
In contrast, $\alpha$--$\beta$ collisions in the complex plane occur for both $n=8$ (b) and $n=20$ (c) somewhere in the range $6<\ell<7$, in good agreement with the estimates $\ell_8\approx 6.03$ and $\ell_{20}\approx 6.84$, respectively.   
Because the $\alpha$ solution has the correct signs for exponential damping when the electron is not recaptured classically, it is the physically correct solution for $\ell>\ell_n$. The $\beta$ solution is an unphysical solution for $\ell>\ell_n$. 

\begin{figure}[ht!]
\centering
    \includegraphics[scale = .35]{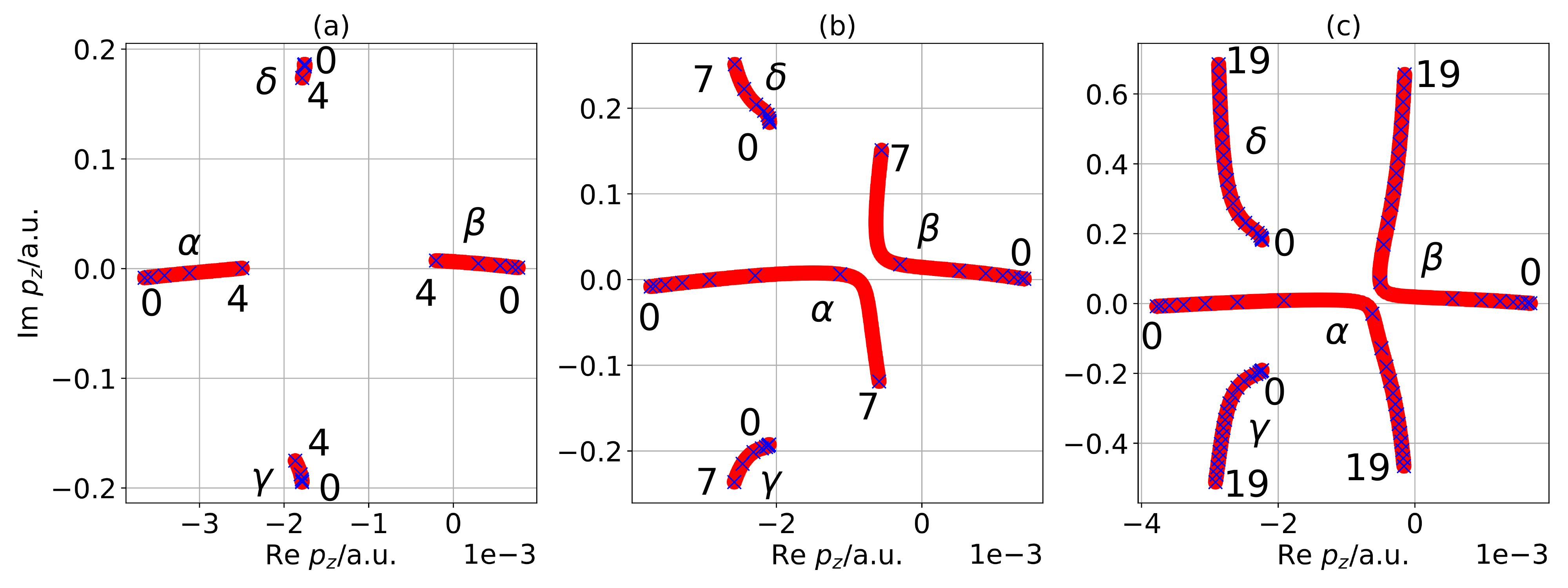}
    \caption{$\ell$ dependence of  $p_z$ for $n=5,8,20$. The numbers in each figure mark the start and end values of $\ell$. The blue crosses mark integer values of $\ell$.}\label{fig:p_z_l_dep}
\end{figure}

While the $\alpha$ and $\beta$ solutions are consistent with the monochromatic field case, discussed in  Ref.~\cite{Popruzhenko_FT}, the two additional solutions, $\gamma$ and $\delta$ in Fig.~\ref{fig:l_dep_cont}, have not been reported previously. 
The $\gamma$ and $\delta$ solutions have completely different behavior as compared with the $\alpha$ and $\beta$ solutions. The real parts of $\gamma$ and $\delta$ are of similar magnitude, while the imaginary parts of $\gamma$ and $\delta$ differ in magnitude. The difference of imaginary parts tends to increase further with $\ell$. Because the imaginary part of $\delta$ is smaller than that of $\alpha$, $\beta$ or $\gamma$ it is essential to know if $\delta$ is a physically sound solution that should be included to model FT from ultrashort laser pulses. 
In order to answer this question, we consider the parallel canonical momentum, $p_z(t_s)$, shown in Fig.~\ref{fig:p_z_l_dep}~(a,b,c) with $t_s$ taken from Fig.~\ref{fig:l_dep_cont}~(a,b,c), respectively. We observe that both the $\alpha$ and $\beta$ solutions have a mostly real $p_z(t_s)$ for low angular momentum, $\ell<\ell_n$. The $\alpha$ solution has a slightly larger magnitude of $\mathrm{Re}[p_z(t_s)]$, which implies a larger final speed of the shorter trajectory ($\alpha$), when compared with the the longer trajectory ($\beta$). Beyond, $\ell>\ell_n$, recapture is not possible and $p_z$, diverge in separate direction along the imaginary axis for the $\alpha$ and $\beta$ solutions. In contrast, the $\gamma$ and $\delta$ solutions have complex $p_z(t_s)$ for all possible angular momenta, which implies that they do not correspond to FT-like solutions, where the electron should evolve in the continuum before the recapture event. We do not have a physical understanding of the $\gamma$ and $\delta$ solutions.

In Fig.~\ref{fig:l_dep_cont}~(d,e,f) we show the imaginary part of the semiclassical action of the solutions, $\mathrm{Im}[S(t_s)]$, computed with the $t_s$ in (a,b,c). This quantity leads to exponential damping of the processes.  In the classically allowed region, $\ell<\ell_n$, we observe that the $\alpha$ and $\beta$ solutions both have approximately the same weight. In the classically forbidden region, the weight of the $\alpha$ solution is reduced with increased angular momentum, $\ell>\ell_n$. In the case of $n=5$ (g) and $n=8$ (h), the weight of $\delta$ is larger than that of the $\gamma$ weight and both $\alpha$ and $\beta$ weights taken in the classical regime. This implies that inclusion of the $\delta$ solution would strongly alter the result of the SFA-FT model.     

In order to ensure that only ``real'' FT solutions are included in the model, we implement a multidimensional filter. This implies that FT trajectories that miss the recapture condition ($\alpha$ and $\beta$ solutions with $\ell>\ell_n$) will be removed, along with all non-FT solutions ($\gamma$, $\delta$, etc solutions). The filter is defined as:
\begin{align}\label{eq:Filter}
\begin{split}
    &\mathrm{Im}\{t_s\} > 0\\
    &\mathrm{Im}\{S_0(t_s)\} < 0\\
    &|\mathrm{Im}\{p_z\}| < 0.1 |\mathrm{Re}\{p_z\}|\\
    &|p_z| < 0.5 \quad \mathrm{ a.u.},
\end{split}
\end{align}
where the first two conditions are standard for saddle point approximations in SFA, and they guarantee that $|\exp (-iS_0)| <1$. The third condition is specific to FT and it selects trajectories that have a classical recapture, while the final condition is used to remove unexpected solutions that originate for a small electric field, i.e. where $E(\mathrm{Re}\{t_s\})$ is close to zero.

\subsubsection{Dependence on carrier--envelope phase of laser pulse}
\begin{figure}[ht!]
\centering
\includegraphics[scale=0.4]{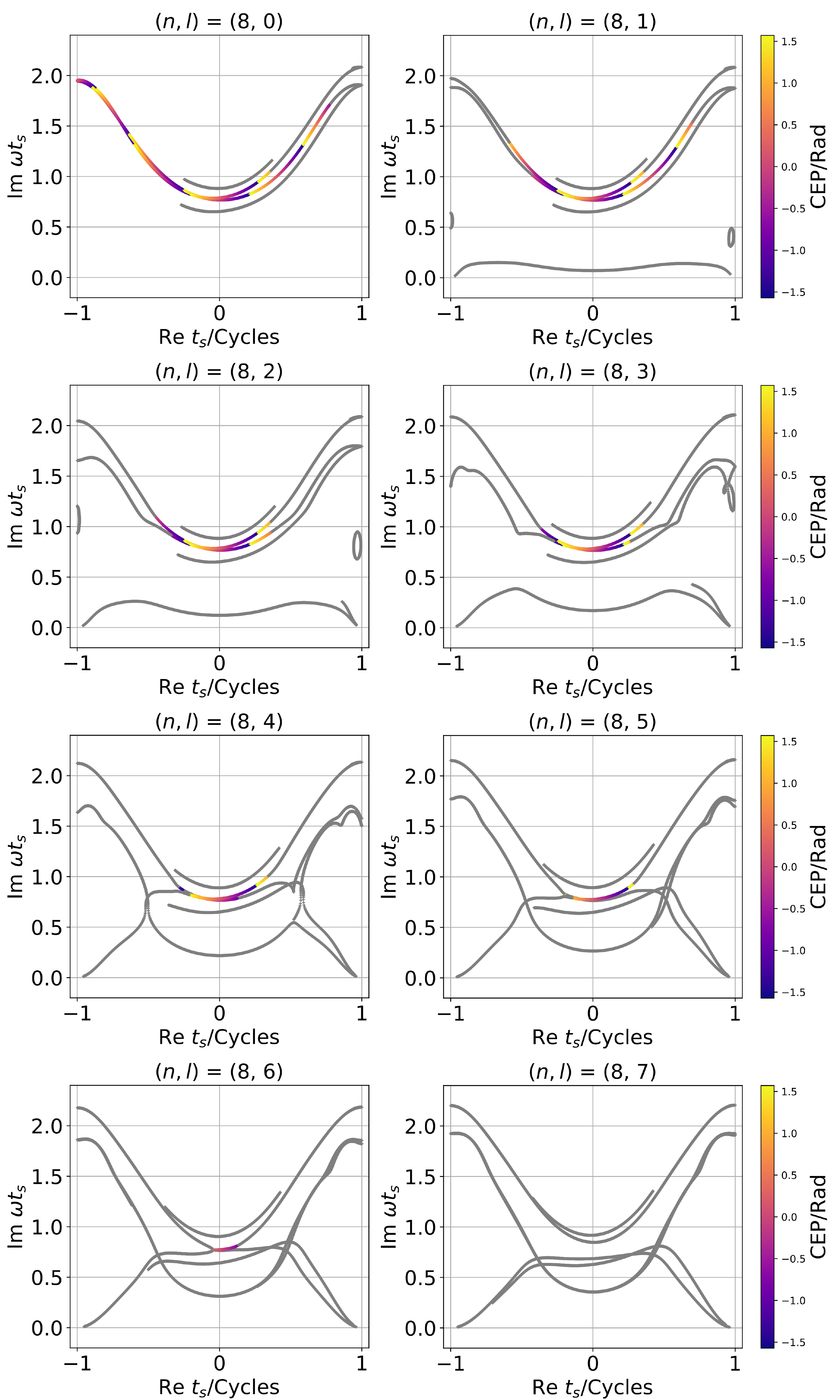}
\caption{CEP and $\ell$ dependence of $t_s$ for $n=8$. The colors belonging to the colorbar to the right of each row indicate which CEP the solutions belong to, while the gray parts of the curves show which solutions are considered to be unphysical.}\label{fig:CEP_dep_t_s} 
\end{figure}

 The CEP of the laser pulse, $\varphi$, is defined in Eq.~(\ref{eq:Action}). Increasing the CEP leads to an advance of the electric field peak. In Fig.~\ref{fig:CEP_dep_t_s}, it is shown how the CEP affects the solutions of Eq. \eqref{eq:EnConReform}, $t_s(\varphi)$. 
 Each individual panel shows $t_s(\varphi)$ for a given $\ell$, ranging from $0$ to $7$, and principal quantum number, $n=8$. All obtained solutions are shown, including the classes: $\alpha$, $\beta$, $\gamma$ and $\delta$, from the previous subsection.
In order to identify the solutions that are real FT solutions, we mark solutions that pass the filter in false CEP color (see colormap). Solutions that are rejected by the filter are drawn with a consistent gray color. 

First, we consider the simplest case, $(n,\ell)=(8,0)$. At the centre of the pulse ($t=0$ cycles), the four solutions: $\alpha$, $\beta$, $\gamma$ and $\delta$ have already been identified. The color scale indicates that for a given CEP, the short trajectory ($\alpha$) is born later than the long trajectory ($\beta$). Physically, this  corresponds to a laser pulse with a electric-field peak with a smaller phase delay. 
The $\alpha$ and $\beta$ solutions can be followed back to the beginning of the pulse ($t=-1$ cycles) by increasing the CEP by more than $\pi$, thus, smoothly connecting classes of solutions in different half cycles of the laser pulse. Decreasing the CEP, which is the same thing as delaying the field peaks, shows that the short trajectory ($\alpha$) stops being a real FT solution rather early in the pulse for $t>0.4$ cycles.  We observe that the long trajectory ($\beta$) remains a real FT solution for an extended time window in the pulse. It develops into non-real FT solutions for $t>0.8$ cycles. In contrast, both $\gamma$ and $\delta$ solutions are non-real FT solutions for all CEP in the entire time window of the pulse.  

When angular momentum is increased, the picture becomes more complicated because additional solutions to Eq.~(\ref{eq:EnConReform}) are found. In the case of $(n,\ell)=(8,1)$ the decreased recapture volume, see Eq.~(\ref{eq:ReturnToRydberg}), leads to $\alpha$ and $\beta$ solutions becoming non-real FT solutions at the start of the pulse. At the centre of pulse both $\alpha$ and $\beta$ are real FT solutions. The start of real FT trajectories on the $\alpha$ solutions is later than that of the $\beta$ solutions.  This indicates that the $\alpha$ trajectories are more sensitive to laser field envelope effects. In addition to the non-real $\gamma$ and $\delta$ solutions, additional solutions are found at the start and end of the laser pulse, as well as at small imaginary time. All of these additional solutions are rejected by the filter as non-real FT solutions and consistently in gray color. 
Further, it can be seen that the filter selects the ``real'' FT solutions, that is to say before the $\alpha$ and $\beta$ solutions split up on the imaginary axis. 
With $(n,\ell)=(8,5)$ we can clearly see the transition from real FT trajectories to the complex case, with the short trajectory ($\alpha$) being the physical solution that decrease exponentially (by increase of its imaginary part), while the long trajectory can be traced to unphysically increased contribution. 
As the angular momentum is increased further, we observe that the time window for real FT trajectories of the $\alpha$ and $\beta$ solutions, is reduced. While a small time window for real FT trajectories remains with $(n,\ell)=8,6$, the highest angular momentum case, $(n,\ell)=(8,7)$, shows that no real FT trajectories are possible for any CEP with the given ultrashort pulse. 

\subsection{Comparison of SFA-FT model with TDSE}
\label{sec:Populations}
\begin{figure}[ht!]
\centering
\includegraphics[scale=.39]{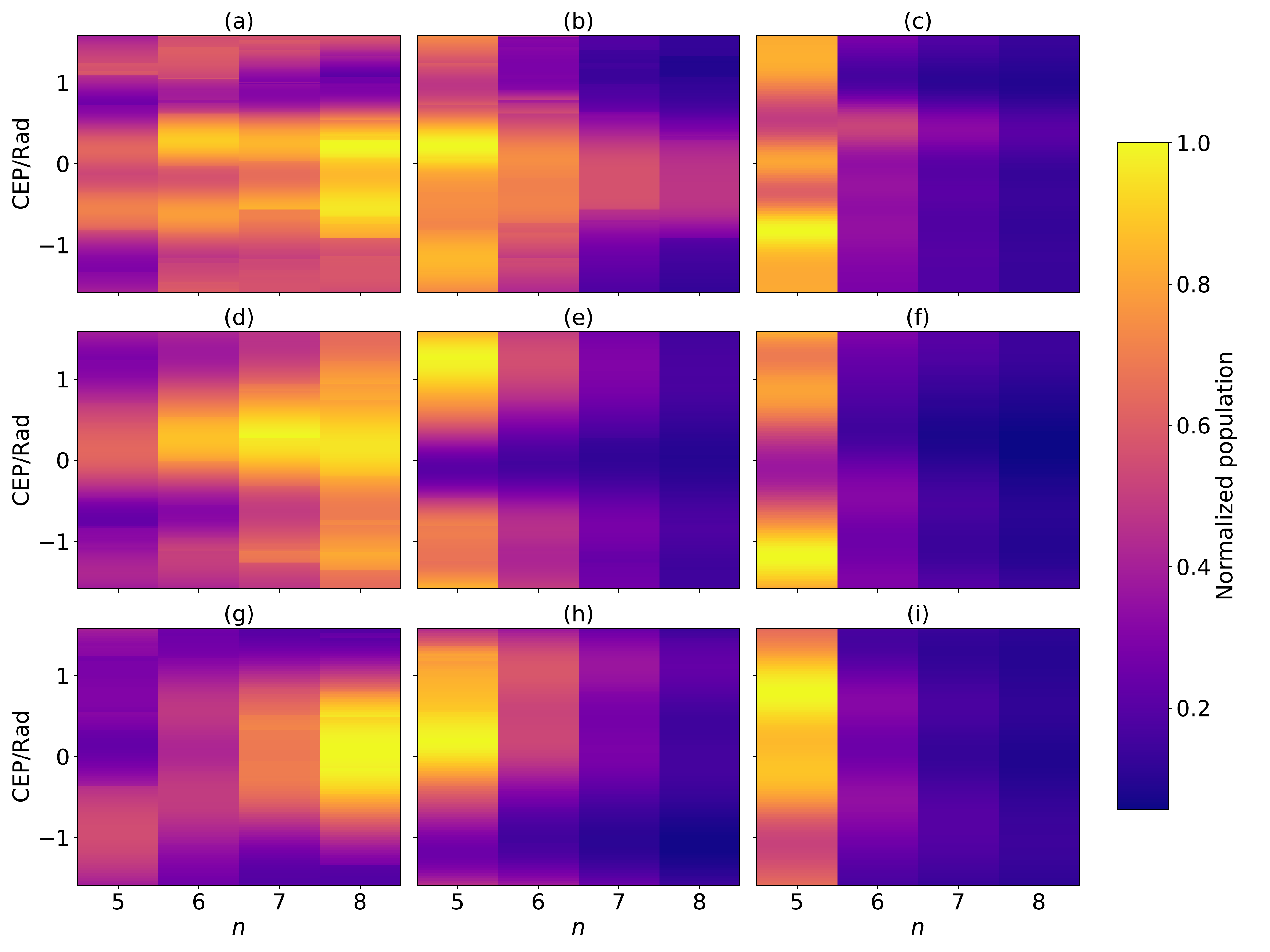}
\caption{CEP dependence of $n$ populations computed with the SFA theory with diffusion (left column), full momentum space wave function prefactor (middle column) and the TDSE (right column)  at the end of  two (top row), three (middle row) and four cycle (bottom row) pulses of \SI{800}{nm} light. The intensity used was $ I = \SI{1.5e14}{W/cm^2}$, which corresponds to a Keldysh parameter of $\gamma = 0.87$. To remove unphysical solutions we used the filter given in Eq. \eqref{eq:Filter}. }\label{fig:Popr_CEP}
\end{figure}

To test if the SFA-FT theory is able to capture the physics that leads to strong field excitation, we present a comparison with results obtained by numerically solving the TDSE. The numerical propagation is done in a spectral basis, with the radial wavefunctions and matrix elements computed by B-splines, and the time propagation is done using a second-order differencing scheme~\cite{Leforestier_1991} in the spectral domain. Propagation has been performed in both length and velocity gauge to ensure that the results are converged and gauge invariant. Figure \ref{fig:Popr_CEP} shows a CEP scan of $n$-populations that was made using the SFA-FT theory and the TDSE. To remove unphysical contributions in the SFA calculations, we use the filter from Eq. \eqref{eq:Filter}.  The left and middle column were computed with the SFA theory, including quantum diffusion, using only the angular part $Y_{\ell m}^{*}(\vecB{p}/\sqrt{p^2})$ (left) or the full $\tilde\phi_{n\ell m}^{*}(\vecB{p})=\left<\phi_{n\ell m}|\vecB{p}\right>$ (middle) for the projection prefactor in Eq.~(\ref{eq:anlmSFA}). The right column shows the results of the TDSE simulations. The top, middle and bottom rows were calculated for  two , three and four cycle pulses, i.e. $\tau=2T$, $3T$ and $4T$ in Eq.~(\ref{eq:Alaserpulse}), respectively.  The populations have been normalized to the maximum of each subplot, so that the same color scale can be used for all plots.  The sharp steps that are seen in the SFA results of Fig. \ref{fig:Popr_CEP}, are due to the filtration process. As shown in Fig. \ref{fig:CEP_dep_t_s}, the evolution of the  SFA-FT solutions is very complicated, which makes filtration  difficult.

As the CEP is varied we see an anti-modulating structure in the TDSE results where a population maximum for one $n=5$ value is accompanied by minima for other $n=6$, and vice versa. This can be clearly seen in Fig.~\ref{fig:Popr_CEP}~(c) at $\varphi = 0.5$. This behavior is not found in the SFA-FT results. When $n \geq 6$, the modulations for different $n$ have similar shapes in both TDSE and SFA. In general the qualitative agreement between SFA-FT and TDSE is better if we use the momentum-space Rydberg state as the projection prefactor, 
$\tilde\phi_{n \ell m}^{*} (\vecB{p})$. 
For instance, the full wavefunction is needed to find the correct decrease in population as $n$ increases. 

\begin{figure}[ht!]
\centering
\includegraphics[scale=.37]{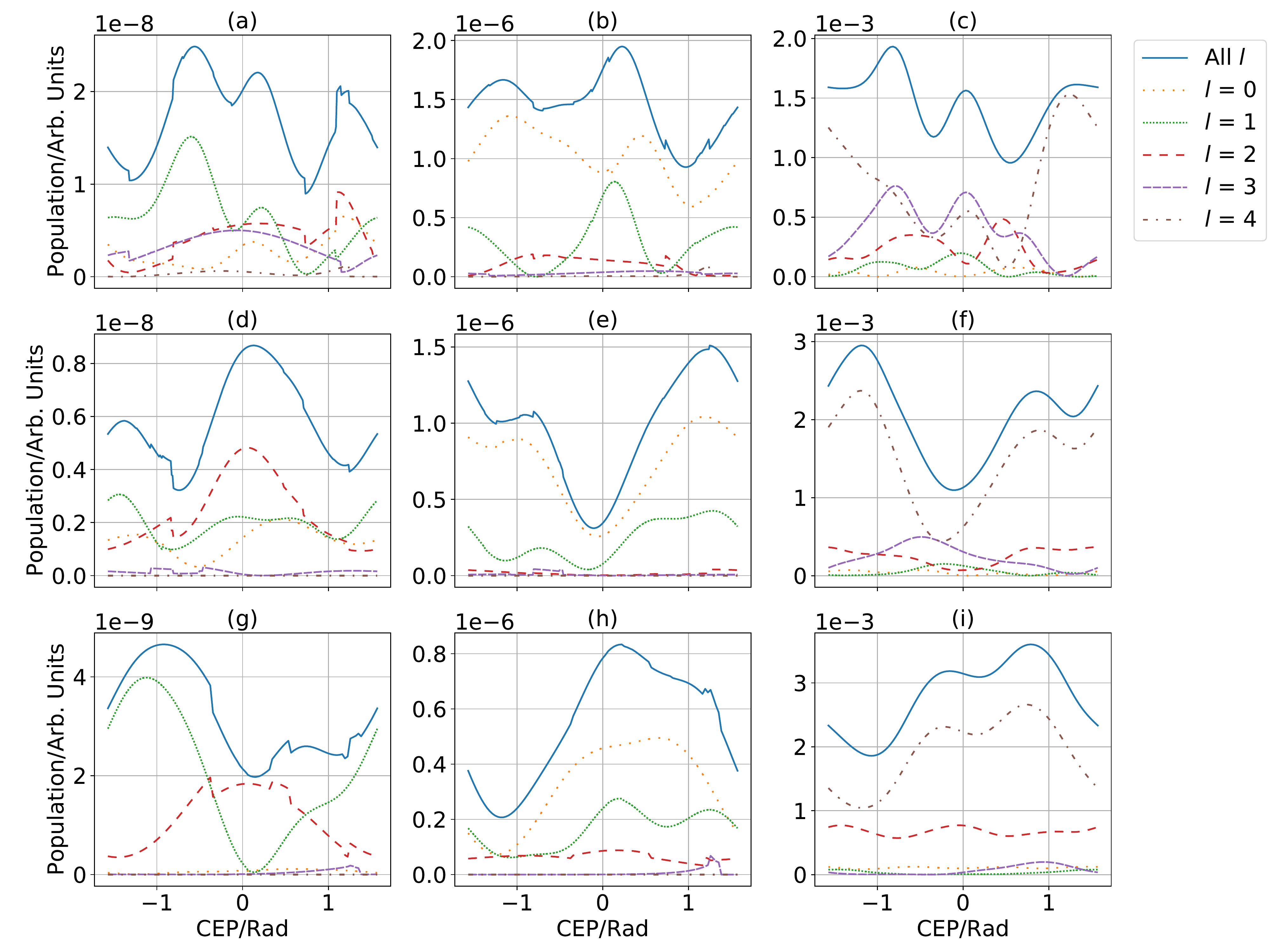}
\caption{Comparison of $\ell$-resolved populations for $n=5$, for two [(a)--(c)], three [(d)--(f)] and four [(g)--(i)] cycle pulses computed with SFA with diffusion (left column), SFA with full momentum space wavefunction (middle column) and TDSE (right column). We used the same pulse parameters as in Fig.~\ref{fig:Popr_CEP}.  }\label{fig:l_resolved_comp}
\end{figure}

Figure~\ref{fig:l_resolved_comp} shows the results for $n=5$ from Fig.~\ref{fig:Popr_CEP}, resolved in $\ell$ and summed over all $\ell$. A qualitative agreement between the total population for $n=5$ is observed between the model (b,e,h) and the TDSE (c,f,i). A maximum is found for a cosine-like pulse (close to $\varphi = 0$) with two and four cycles, while a minimum is found for three cycles. The strength of the peaks as a function of CEP are not correctly described by the model, and it is found that the projection factor is essential for the qualitative agreement with TDSE, see Fig.~\ref{fig:l_resolved_comp} (a,d,g). The fact that we see several maxima as a function of CEP, is a sign of quantum interference between different trajectories.  In contrast, if the contributions from different trajectories are summed incoherently a single smooth maximum is observed in the given CEP range, as expected from the  tunneling step of the process (not shown). From these plots it is clear that although we can find qualitative agreement between the SFA-FT and TDSE results when the results are summed over $\ell$, the contributions from individual $\ell$ channels are completely different in the two approaches. In the TDSE results, the total population for $n=5$ comes mostly from $\ell\geq2$, with a tendency towards being completely dominated by $\ell=4$ for the longer pulses. For both SFA calculations we mostly see contributions from $\ell\leq 2$, with $\ell=4$ barely contributing. There is also a difference between the two SFA prefactors. When we only use the angular part of the momentum space wavefunction, the dominant contributions come from $\ell=1$ and $\ell=2$, while when we include the full wavefunction the dominant contributions come from $\ell=0$ and $\ell=1$. This can explain why we see larger discontinuities for the total population in $n=5$ when we only include the angular factor, since more solutions are filtered out as $\ell$ increases, see Fig.~\ref{fig:CEP_dep_t_s}. Excitation to $n=5$ corresponds to 8.42 laser photons, which is in between even and odd photon numbers. With ultrashort pulses the width of the photons allows for overlapping parity to the final state. For the shortest pulses considered, even and odd parity states have comparable populations, see Fig.~\ref{fig:l_resolved_comp}~(c), while for the longest pulses considered, even parity is favored, see Fig.~\ref{fig:l_resolved_comp}~(i). Similar mixing of parity is observed also in the SFA-FT model.

\begin{figure}[ht!]
\centering
\includegraphics[scale=.45]{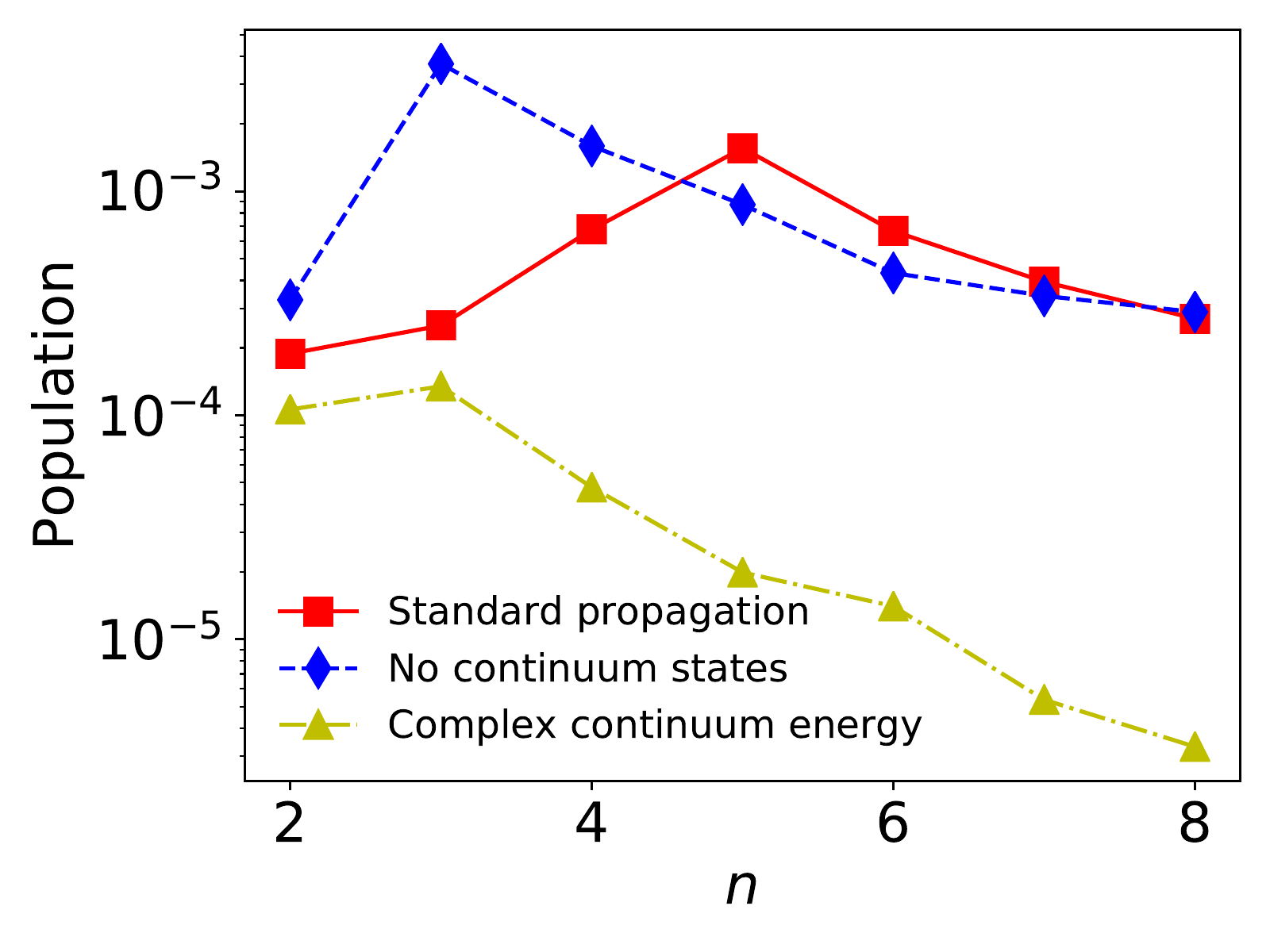}
\caption{Population distribution in $n$ for standard TDSE propagation, restricting the state space to bound states, and when adding an imaginary part of $-0.5$~Ha to the energy of the continuum states.}\label{fig:Cont_damp}
\end{figure}

In order to better understand the role that the continuum states play in the strong-field population of excited states, we performed TDSE simulations where we either remove them entirely from the state space, or add an imaginary part to the energy of each continuum state so that their amplitudes become damped over a subcycle timescale during propagation. These calculations were performed in length gauge. We found that the continuum states play a crucial role in getting the correct excited state distribution, and it appears to be especially important for $n\geq 4$. Figure \ref{fig:Cont_damp} shows a comparison between the results of standard propagation, removing the continuum states, and changing the energy of the continuum states to $E \rightarrow E -0.5 i$. The results for $n=2$ and $n=3$ have not been affected very much by the addition of complex continuum energies, whereas the higher excited states show a significant difference to the results of the standard propagation. Propagating without the continuum states predicts a maximum at $n=3$ instead of $n=5$, but the yield to higher $n$ is similar. This suggests that the FT model can have qualitative agreement with the TDSE results for $n\geq 4$, and why the same is not true for $n=2,3$. We find similar results when the the imaginary part of the energy is $1$ and $0.1$.

\section{Conclusion}\label{sec:conclusion}
We have studied FT in hydrogen with ultrashort laser pulses,  using the SFA-based theory presented in Ref.~\cite{Popruzhenko_FT}. Specifically we studied the effects the CEP on the population of excited states. Comparison with results from the TDSE shows that there is qualitative agreement in the CEP modulations when the populations are summed over $\ell$, but that the contributions from individual $\ell$ channels do not agree. 
As already mentioned in Ref.~\cite{Popruzhenko_FT}, SFA trajectories constitute a rough approximation  to the exact electron dynamics, and the theory is therefore unable to capture Coulomb effects during propagation. We find that there are numerous solutions to the SFA model within the time window of an ultrashort laser pulse. Some of these solutions  correspond to FT processes, and should be added coherently, but other solutions are unphysical and must be excluded by a complex filtering process. Two physical solutions per electric field peak, denoted $\alpha$ and $\beta$, are found at low angular momenta in agreement with Ref.~\cite{Popruzhenko_FT}. We find that the $\alpha$ solution is born after the field peak, while the $\beta$ solution is born before the field peak. This implies that the $\alpha$ solutions are colliding trajectories with the ion, similar to the short and long trajectories in HHG, while the $\beta$ solutions is a different class of trajectories that drift away from the atom. In future work, it may be possible to treat the different parts of the process by more exact methods, improving the description of the trajectories and action \cite{Smirnova_Ivanov_book,Smirnova2008,Popruzhenko_2014}. This would be especially interesting for the $\beta$ solutions that do not return to the Coulomb singularity. We may speculate that the $\beta$ solutions are the physically most significant trajectories because the $\alpha$ solutions would be strongly scattered by the Coulomb potential before the end of the laser pulse in a more exact treatment. This prediction is well supported by classical simulations, where it has been found that trajectories that are born before the peak of the laser oscillations lead to the dominant recapture process on Rydberg states
\cite{Shvetsov_2009,Eichmann_2016,Dubois2018,Zhao2019}. At high angular momenta we find no physical solutions to the SFA model, because the capture volume becomes smaller, and the momenta of both $\alpha$ and $\beta$ solutions become complex. This trend of low probability for high $\ell$ states is in disagreement with our exact TDSE results where we see significant contributions from $\ell$ up to about $\ell = 7$, which agrees with the estimate given in Sec.~\ref{sec:interactiontime}. Classical ensemble simulations show higher angular momentum distributions, suggesting that Coulomb effects are responsible for the redistribution in angular momentum \cite{Zhao2019}. 

There is room for future improvements of the considered SFA-FT model because the saddle-point equation is not derived from consistent first principles, e.g. as a Dyson expansion. Indeed, an alternative SFA-FT theory was recently proposed based on a second order Dyson expansion \cite{Hu:19}. To make the models more consistent it would be satisfactory if the saddle-point equations could be derived as saddle-point equations of an appropriate action, as is done in SFA for HHG~\cite{HHG_Lewenstein}. However, as we have shown, a straightforward application of the saddle-point method to the SFA action only amounts to quantum diffusion and trajectories that return to the origin. While we find that quantum diffusion affects the relative strength of contributions within the laser pulse, it did not modify the structure populations over the CEP effects in a significant way with the considered laser parameters. 
Finally, we have performed TDSE simulations to study the role of the continuum in FT from ultrashort laser pulses and it is found that a correct treatment of continuum states is important to find amplitudes for the highly excited Rydberg states states, thus, giving confidence that FT is a meaningful physical description of the ultra-fast Rydberg excitation process.   

\begin{acknowledgements}
JMD acknowledges support from the Swedish Research Council: 2018-03845, the Olle Engkvist Foundation: 194-0734 and the Knut and Alice Wallenberg Foundation: 2017.0104 and 2019.0154. We also acknowledge stimulating discussions with Emilio Pisanty.
\end{acknowledgements}

\section*{References}
\providecommand{\noopsort}[1]{}\providecommand{\singleletter}[1]{#1}%

\begin{thebibliography}{23}%
\makeatletter
\providecommand \@ifxundefined [1]{%
 \@ifx{#1\undefined}
}%
\providecommand \@ifnum [1]{%
 \ifnum #1\expandafter \@firstoftwo
 \else \expandafter \@secondoftwo
 \fi
}%
\providecommand \@ifx [1]{%
 \ifx #1\expandafter \@firstoftwo
 \else \expandafter \@secondoftwo
 \fi
}%
\providecommand \natexlab [1]{#1}%
\providecommand \enquote  [1]{``#1''}%
\providecommand \bibnamefont  [1]{#1}%
\providecommand \bibfnamefont [1]{#1}%
\providecommand \citenamefont [1]{#1}%
\providecommand \href@noop [0]{\@secondoftwo}%
\providecommand \href [0]{\begingroup \@sanitize@url \@href}%
\providecommand \@href[1]{\@@startlink{#1}\@@href}%
\providecommand \@@href[1]{\endgroup#1\@@endlink}%
\providecommand \@sanitize@url [0]{\catcode `\\12\catcode `\$12\catcode
  `\&12\catcode `\#12\catcode `\^12\catcode `\_12\catcode `\%12\relax}%
\providecommand \@@startlink[1]{}%
\providecommand \@@endlink[0]{}%
\providecommand \url  [0]{\begingroup\@sanitize@url \@url }%
\providecommand \@url [1]{\endgroup\@href {#1}{\urlprefix }}%
\providecommand \urlprefix  [0]{URL }%
\providecommand \Eprint [0]{\href }%
\providecommand \doibase [0]{https://doi.org/}%
\providecommand \selectlanguage [0]{\@gobble}%
\providecommand \bibinfo  [0]{\@secondoftwo}%
\providecommand \bibfield  [0]{\@secondoftwo}%
\providecommand \translation [1]{[#1]}%
\providecommand \BibitemOpen [0]{}%
\providecommand \bibitemStop [0]{}%
\providecommand \bibitemNoStop [0]{.\EOS\space}%
\providecommand \EOS [0]{\spacefactor3000\relax}%
\providecommand \BibitemShut  [1]{\csname bibitem#1\endcsname}%
\let\auto@bib@innerbib\@empty
%</preamble>
\bibitem [{\citenamefont {Nubbemeyer}\ \emph {et~al.}(2008)\citenamefont
  {Nubbemeyer}, \citenamefont {Gorling}, \citenamefont {Saenz}, \citenamefont
  {Eichmann},\ and\ \citenamefont {Sandner}}]{FT_original}%
  \BibitemOpen
  \bibfield  {author} {\bibinfo {author} {\bibfnamefont {T.}~\bibnamefont
  {Nubbemeyer}}, \bibinfo {author} {\bibfnamefont {K.}~\bibnamefont {Gorling}},
  \bibinfo {author} {\bibfnamefont {A.}~\bibnamefont {Saenz}}, \bibinfo
  {author} {\bibfnamefont {U.}~\bibnamefont {Eichmann}},\ and\ \bibinfo
  {author} {\bibfnamefont {W.}~\bibnamefont {Sandner}},\ }\href
  {https://doi.org/10.1103/PhysRevLett.101.233001} {\bibfield  {journal}
  {\bibinfo  {journal} {Phys. Rev. Lett.}\ }\textbf {\bibinfo {volume} {101}},\
  \bibinfo {pages} {233001} (\bibinfo {year} {2008})}\BibitemShut {NoStop}%
\bibitem [{\citenamefont {Gavrila}(2002)}]{Gavrila2002}%
  \BibitemOpen
  \bibfield  {author} {\bibinfo {author} {\bibfnamefont {M.}~\bibnamefont
  {Gavrila}},\ }\href {https://doi.org/10.1088/0953-4075/35/18/201} {\bibfield
  {journal} {\bibinfo  {journal} {Journal of Physics B: Atomic, Molecular and
  Optical Physics}\ }\textbf {\bibinfo {volume} {35}},\ \bibinfo {pages} {R147}
  (\bibinfo {year} {2002})}\BibitemShut {NoStop}%
\bibitem [{\citenamefont {Eichmann}\ \emph {et~al.}(2009)\citenamefont
  {Eichmann}, \citenamefont {Nubbemeyer}, \citenamefont {Rottke},\ and\
  \citenamefont {Sandner}}]{Eichmann2009}%
  \BibitemOpen
  \bibfield  {author} {\bibinfo {author} {\bibfnamefont {U.}~\bibnamefont
  {Eichmann}}, \bibinfo {author} {\bibfnamefont {T.}~\bibnamefont
  {Nubbemeyer}}, \bibinfo {author} {\bibfnamefont {H.}~\bibnamefont {Rottke}},\
  and\ \bibinfo {author} {\bibfnamefont {W.}~\bibnamefont {Sandner}},\ }\href
  {https://doi.org/10.1038/nature08481} {\bibfield  {journal} {\bibinfo
  {journal} {Nature}\ }\textbf {\bibinfo {volume} {461}},\ \bibinfo {pages}
  {1261} (\bibinfo {year} {2009})}\BibitemShut {NoStop}%
\bibitem [{\citenamefont {Morales}\ \emph {et~al.}(2011)\citenamefont
  {Morales}, \citenamefont {Richter}, \citenamefont {Patchkovskii},\ and\
  \citenamefont {Smirnova}}]{Morales2011}%
  \BibitemOpen
  \bibfield  {author} {\bibinfo {author} {\bibfnamefont {F.}~\bibnamefont
  {Morales}}, \bibinfo {author} {\bibfnamefont {M.}~\bibnamefont {Richter}},
  \bibinfo {author} {\bibfnamefont {S.}~\bibnamefont {Patchkovskii}},\ and\
  \bibinfo {author} {\bibfnamefont {O.}~\bibnamefont {Smirnova}},\ }\href
  {https://doi.org/10.1073/pnas.1105916108} {\bibfield  {journal} {\bibinfo
  {journal} {Proceedings of the National Academy of Sciences}\ }\textbf
  {\bibinfo {volume} {108}},\ \bibinfo {pages} {16906} (\bibinfo {year}
  {2011})}\BibitemShut {NoStop}%
\bibitem [{\citenamefont {Kjellsson}\ \emph {et~al.}(2017)\citenamefont
  {Kjellsson}, \citenamefont {Selst{\o}},\ and\ \citenamefont
  {Lindroth}}]{Kjellsson2017b}%
  \BibitemOpen
  \bibfield  {author} {\bibinfo {author} {\bibfnamefont {T.}~\bibnamefont
  {Kjellsson}}, \bibinfo {author} {\bibfnamefont {S.}~\bibnamefont
  {Selst{\o}}},\ and\ \bibinfo {author} {\bibfnamefont {E.}~\bibnamefont
  {Lindroth}},\ }\href {https://doi.org/10.1103/physreva.95.043403} {\bibfield
  {journal} {\bibinfo  {journal} {Physical Review A}\ }\textbf {\bibinfo
  {volume} {95}},\ \bibinfo {pages} {043403} (\bibinfo {year}
  {2017})}\BibitemShut {NoStop}%
\bibitem [{\citenamefont {Shvetsov-Shilovski}\ \emph
  {et~al.}(2009)\citenamefont {Shvetsov-Shilovski}, \citenamefont
  {Goreslavski}, \citenamefont {Popruzhenko},\ and\ \citenamefont
  {Becker}}]{Shvetsov_2009}%
  \BibitemOpen
  \bibfield  {author} {\bibinfo {author} {\bibfnamefont {N.~I.}\ \bibnamefont
  {Shvetsov-Shilovski}}, \bibinfo {author} {\bibfnamefont {S.~P.}\ \bibnamefont
  {Goreslavski}}, \bibinfo {author} {\bibfnamefont {S.~V.}\ \bibnamefont
  {Popruzhenko}},\ and\ \bibinfo {author} {\bibfnamefont {W.}~\bibnamefont
  {Becker}},\ }\href {https://doi.org/10.1134/S1054660X09150377} {\bibfield
  {journal} {\bibinfo  {journal} {Laser Phys.}\ }\textbf {\bibinfo {volume}
  {19}},\ \bibinfo {pages} {1550} (\bibinfo {year} {2009})}\BibitemShut
  {NoStop}%
\bibitem [{\citenamefont {de~Boer}\ and\ \citenamefont
  {Muller}(1992)}]{Muller_1992}%
  \BibitemOpen
  \bibfield  {author} {\bibinfo {author} {\bibfnamefont {M.~P.}\ \bibnamefont
  {de~Boer}}\ and\ \bibinfo {author} {\bibfnamefont {H.~G.}\ \bibnamefont
  {Muller}},\ }\href {https://doi.org/10.1103/PhysRevLett.68.2747} {\bibfield
  {journal} {\bibinfo  {journal} {Phys. Rev. Lett.}\ }\textbf {\bibinfo
  {volume} {68}},\ \bibinfo {pages} {2747} (\bibinfo {year}
  {1992})}\BibitemShut {NoStop}%
\bibitem [{\citenamefont {Zimmermann}\ \emph {et~al.}(2017)\citenamefont
  {Zimmermann}, \citenamefont {Patchkovskii}, \citenamefont {Ivanov},\ and\
  \citenamefont {Eichmann}}]{Zimmerman_2017}%
  \BibitemOpen
  \bibfield  {author} {\bibinfo {author} {\bibfnamefont {H.}~\bibnamefont
  {Zimmermann}}, \bibinfo {author} {\bibfnamefont {S.}~\bibnamefont
  {Patchkovskii}}, \bibinfo {author} {\bibfnamefont {M.~Y.}\ \bibnamefont
  {Ivanov}},\ and\ \bibinfo {author} {\bibfnamefont {U.}~\bibnamefont
  {Eichmann}},\ }\href {https://doi.org/10.1103/PhysRevLett.118.013003}
  {\bibfield  {journal} {\bibinfo  {journal} {Phys. Rev. Lett.}\ }\textbf
  {\bibinfo {volume} {118}},\ \bibinfo {pages} {013003} (\bibinfo {year}
  {2017})}\BibitemShut {NoStop}%
\bibitem [{\citenamefont {Chen}\ \emph {et~al.}(2012)\citenamefont {Chen},
  \citenamefont {Gao}, \citenamefont {Li}, \citenamefont {Becker},\ and\
  \citenamefont {Jaro\'{n}-Becker}}]{Chen_2012}%
  \BibitemOpen
  \bibfield  {author} {\bibinfo {author} {\bibfnamefont {S.}~\bibnamefont
  {Chen}}, \bibinfo {author} {\bibfnamefont {X.}~\bibnamefont {Gao}}, \bibinfo
  {author} {\bibfnamefont {J.}~\bibnamefont {Li}}, \bibinfo {author}
  {\bibfnamefont {A.}~\bibnamefont {Becker}},\ and\ \bibinfo {author}
  {\bibfnamefont {A.}~\bibnamefont {Jaro\'{n}-Becker}},\ }\href
  {https://doi.org/10.1103/PhysRevA.86.013410} {\bibfield  {journal} {\bibinfo
  {journal} {Phys. Rev. A}\ }\textbf {\bibinfo {volume} {86}},\ \bibinfo
  {pages} {013410} (\bibinfo {year} {2012})}\BibitemShut {NoStop}%
\bibitem [{\citenamefont {Lewenstein}\ \emph {et~al.}(1994)\citenamefont
  {Lewenstein}, \citenamefont {Balcou}, \citenamefont {Ivanov}, \citenamefont
  {L'Huillier},\ and\ \citenamefont {Corkum}}]{HHG_Lewenstein}%
  \BibitemOpen
  \bibfield  {author} {\bibinfo {author} {\bibfnamefont {M.}~\bibnamefont
  {Lewenstein}}, \bibinfo {author} {\bibfnamefont {P.}~\bibnamefont {Balcou}},
  \bibinfo {author} {\bibfnamefont {M.~Y.}\ \bibnamefont {Ivanov}}, \bibinfo
  {author} {\bibfnamefont {A.}~\bibnamefont {L'Huillier}},\ and\ \bibinfo
  {author} {\bibfnamefont {P.~B.}\ \bibnamefont {Corkum}},\ }\href
  {https://doi.org/10.1103/PhysRevA.49.2117} {\bibfield  {journal} {\bibinfo
  {journal} {Phys. Rev. A}\ }\textbf {\bibinfo {volume} {49}},\ \bibinfo
  {pages} {2117} (\bibinfo {year} {1994})}\BibitemShut {NoStop}%
\bibitem [{\citenamefont {Becker}\ \emph {et~al.}(2002)\citenamefont {Becker},
  \citenamefont {Grasbon}, \citenamefont {Kopold}, \citenamefont
  {Milo\v{s}evi\'{c}}, \citenamefont {Paulus},\ and\ \citenamefont
  {Walther}}]{Becker_2002}%
  \BibitemOpen
  \bibfield  {author} {\bibinfo {author} {\bibfnamefont {W.}~\bibnamefont
  {Becker}}, \bibinfo {author} {\bibfnamefont {F.}~\bibnamefont {Grasbon}},
  \bibinfo {author} {\bibfnamefont {R.}~\bibnamefont {Kopold}}, \bibinfo
  {author} {\bibfnamefont {D.}~\bibnamefont {Milo\v{s}evi\'{c}}}, \bibinfo
  {author} {\bibfnamefont {G.}~\bibnamefont {Paulus}},\ and\ \bibinfo {author}
  {\bibfnamefont {H.}~\bibnamefont {Walther}}\ }(\bibinfo  {publisher}
  {Academic Press},\ \bibinfo {year} {2002})\ pp.\ \bibinfo {pages}
  {35--98}\BibitemShut {NoStop}%
\bibitem [{\citenamefont {Popruzhenko}(2017)}]{Popruzhenko_FT}%
  \BibitemOpen
  \bibfield  {author} {\bibinfo {author} {\bibfnamefont {S.~V.}\ \bibnamefont
  {Popruzhenko}},\ }\href {https://doi.org/10.1088/1361-6455/aa948b} {\bibfield
   {journal} {\bibinfo  {journal} {J. Phys. B: At. Mol. Opt. Phys.}\ }\textbf
  {\bibinfo {volume} {51}},\ \bibinfo {pages} {014002} (\bibinfo {year}
  {2017})}\BibitemShut {NoStop}%
\bibitem [{\citenamefont {Hu}\ \emph {et~al.}(2019)\citenamefont {Hu},
  \citenamefont {Hao}, \citenamefont {Lv}, \citenamefont {Liu}, \citenamefont
  {Yang}, \citenamefont {Xu}, \citenamefont {Jin}, \citenamefont {Ding},
  \citenamefont {Li}, \citenamefont {Li}, \citenamefont {Becker},\ and\
  \citenamefont {Chen}}]{Hu:19}%
  \BibitemOpen
  \bibfield  {author} {\bibinfo {author} {\bibfnamefont {S.}~\bibnamefont
  {Hu}}, \bibinfo {author} {\bibfnamefont {X.}~\bibnamefont {Hao}}, \bibinfo
  {author} {\bibfnamefont {H.}~\bibnamefont {Lv}}, \bibinfo {author}
  {\bibfnamefont {M.}~\bibnamefont {Liu}}, \bibinfo {author} {\bibfnamefont
  {T.}~\bibnamefont {Yang}}, \bibinfo {author} {\bibfnamefont {H.}~\bibnamefont
  {Xu}}, \bibinfo {author} {\bibfnamefont {M.}~\bibnamefont {Jin}}, \bibinfo
  {author} {\bibfnamefont {D.}~\bibnamefont {Ding}}, \bibinfo {author}
  {\bibfnamefont {Q.}~\bibnamefont {Li}}, \bibinfo {author} {\bibfnamefont
  {W.}~\bibnamefont {Li}}, \bibinfo {author} {\bibfnamefont {W.}~\bibnamefont
  {Becker}},\ and\ \bibinfo {author} {\bibfnamefont {J.}~\bibnamefont {Chen}},\
  }\href {https://doi.org/10.1364/OE.27.031629} {\bibfield  {journal} {\bibinfo
   {journal} {Opt. Express}\ }\textbf {\bibinfo {volume} {27}},\ \bibinfo
  {pages} {31629} (\bibinfo {year} {2019})}\BibitemShut {NoStop}%
\bibitem [{\citenamefont {Nakajima}\ and\ \citenamefont
  {Watanabe}(2006{\natexlab{a}})}]{Nakajima_2006_1}%
  \BibitemOpen
  \bibfield  {author} {\bibinfo {author} {\bibfnamefont {T.}~\bibnamefont
  {Nakajima}}\ and\ \bibinfo {author} {\bibfnamefont {S.}~\bibnamefont
  {Watanabe}},\ }\href {https://doi.org/10.1103/PhysRevLett.96.213001}
  {\bibfield  {journal} {\bibinfo  {journal} {Phys. Rev. Lett.}\ }\textbf
  {\bibinfo {volume} {96}},\ \bibinfo {pages} {213001} (\bibinfo {year}
  {2006}{\natexlab{a}})}\BibitemShut {NoStop}%
\bibitem [{\citenamefont {Nakajima}\ and\ \citenamefont
  {Watanabe}(2006{\natexlab{b}})}]{Nakajima_2006_2}%
  \BibitemOpen
  \bibfield  {author} {\bibinfo {author} {\bibfnamefont {T.}~\bibnamefont
  {Nakajima}}\ and\ \bibinfo {author} {\bibfnamefont {S.}~\bibnamefont
  {Watanabe}},\ }\href {https://doi.org/10.1364/OL.31.001920} {\bibfield
  {journal} {\bibinfo  {journal} {Opt. Lett.}\ }\textbf {\bibinfo {volume}
  {31}},\ \bibinfo {pages} {1920} (\bibinfo {year}
  {2006}{\natexlab{b}})}\BibitemShut {NoStop}%
\bibitem [{\citenamefont {Smirnova}\ and\ \citenamefont
  {Ivanov}(2014)}]{Smirnova_Ivanov_book}%
  \BibitemOpen
  \bibfield  {author} {\bibinfo {author} {\bibfnamefont {O.}~\bibnamefont
  {Smirnova}}\ and\ \bibinfo {author} {\bibfnamefont {M.~Y.}\ \bibnamefont
  {Ivanov}},\ }\bibinfo {title} {Multielectron high harmonic generation: Simple
  man on a complex plane},\ in\ \href
  {https://doi.org/https://doi.org/10.1002/9783527677689.ch7} {\emph {\bibinfo
  {booktitle} {Attosecond and XUV Physics}}}\ (\bibinfo  {publisher} {John
  Wiley \& Sons, Ltd},\ \bibinfo {year} {2014})\ Chap.~\bibinfo {chapter} {7},
  pp.\ \bibinfo {pages} {201--256},\ \Eprint
  {https://arxiv.org/abs/https://onlinelibrary.wiley.com/doi/pdf/10.1002/9783527677689.ch7}
  {https://onlinelibrary.wiley.com/doi/pdf/10.1002/9783527677689.ch7}
  \BibitemShut {NoStop}%
\bibitem [{\citenamefont {Bethe}\ and\ \citenamefont
  {Salpeter}(1977)}]{Bethe1977}%
  \BibitemOpen
  \bibfield  {author} {\bibinfo {author} {\bibfnamefont {H.}~\bibnamefont
  {Bethe}}\ and\ \bibinfo {author} {\bibfnamefont {E.}~\bibnamefont
  {Salpeter}},\ }\href@noop {} {\emph {\bibinfo {title} {Quantum mechanics of
  one- and two-electron atoms}}}\ (\bibinfo  {publisher} {Plenum Pub. Corp},\
  \bibinfo {address} {New York},\ \bibinfo {year} {1977})\BibitemShut {NoStop}%
\bibitem [{\citenamefont {Leforestier}\ \emph {et~al.}(1991)\citenamefont
  {Leforestier}, \citenamefont {Bisseling}, \citenamefont {Cerjan},
  \citenamefont {Feit}, \citenamefont {Friesner}, \citenamefont {Guldberg},
  \citenamefont {Hammerich}, \citenamefont {Jolicard}, \citenamefont
  {Karrlein}, \citenamefont {Meyer}, \citenamefont {Lipkin}, \citenamefont
  {Roncero},\ and\ \citenamefont {Kosloff}}]{Leforestier_1991}%
  \BibitemOpen
  \bibfield  {author} {\bibinfo {author} {\bibfnamefont {C.}~\bibnamefont
  {Leforestier}}, \bibinfo {author} {\bibfnamefont {R.}~\bibnamefont
  {Bisseling}}, \bibinfo {author} {\bibfnamefont {C.}~\bibnamefont {Cerjan}},
  \bibinfo {author} {\bibfnamefont {M.}~\bibnamefont {Feit}}, \bibinfo {author}
  {\bibfnamefont {R.}~\bibnamefont {Friesner}}, \bibinfo {author}
  {\bibfnamefont {A.}~\bibnamefont {Guldberg}}, \bibinfo {author}
  {\bibfnamefont {A.}~\bibnamefont {Hammerich}}, \bibinfo {author}
  {\bibfnamefont {G.}~\bibnamefont {Jolicard}}, \bibinfo {author}
  {\bibfnamefont {W.}~\bibnamefont {Karrlein}}, \bibinfo {author}
  {\bibfnamefont {H.-D.}\ \bibnamefont {Meyer}}, \bibinfo {author}
  {\bibfnamefont {N.}~\bibnamefont {Lipkin}}, \bibinfo {author} {\bibfnamefont
  {O.}~\bibnamefont {Roncero}},\ and\ \bibinfo {author} {\bibfnamefont
  {R.}~\bibnamefont {Kosloff}},\ }\href
  {https://doi.org/https://doi.org/10.1016/0021-9991(91)90137-A} {\bibfield
  {journal} {\bibinfo  {journal} {Journal of Computational Physics}\ }\textbf
  {\bibinfo {volume} {94}},\ \bibinfo {pages} {59} (\bibinfo {year}
  {1991})}\BibitemShut {NoStop}%
\bibitem [{\citenamefont {Smirnova}\ \emph {et~al.}(2008)\citenamefont
  {Smirnova}, \citenamefont {Spanner},\ and\ \citenamefont
  {Ivanov}}]{Smirnova2008}%
  \BibitemOpen
  \bibfield  {author} {\bibinfo {author} {\bibfnamefont {O.}~\bibnamefont
  {Smirnova}}, \bibinfo {author} {\bibfnamefont {M.}~\bibnamefont {Spanner}},\
  and\ \bibinfo {author} {\bibfnamefont {M.~Y.}\ \bibnamefont {Ivanov}},\
  }\bibfield  {journal} {\bibinfo  {journal} {Physical Review A}\ }\textbf
  {\bibinfo {volume} {77}},\ \href {https://doi.org/10.1103/physreva.77.033407}
  {10.1103/physreva.77.033407} (\bibinfo {year} {2008})\BibitemShut {NoStop}%
\bibitem [{\citenamefont {Popruzhenko}(2014)}]{Popruzhenko_2014}%
  \BibitemOpen
  \bibfield  {author} {\bibinfo {author} {\bibfnamefont {S.~V.}\ \bibnamefont
  {Popruzhenko}},\ }\href {https://doi.org/10.1088/0953-4075/47/20/204001}
  {\bibfield  {journal} {\bibinfo  {journal} {Journal of Physics B: Atomic,
  Molecular and Optical Physics}\ }\textbf {\bibinfo {volume} {47}},\ \bibinfo
  {pages} {204001} (\bibinfo {year} {2014})}\BibitemShut {NoStop}%
\bibitem [{\citenamefont {Eichmann}(2016)}]{Eichmann_2016}%
  \BibitemOpen
  \bibfield  {author} {\bibinfo {author} {\bibfnamefont {U.}~\bibnamefont
  {Eichmann}},\ }\bibinfo {title} {Strong-field induced atomic excitation and
  kinematics},\ in\ \href {https://doi.org/10.1007/978-3-319-20173-3_1} {\emph
  {\bibinfo {booktitle} {Ultrafast Dynamics Driven by Intense Light Pulses:
  From Atoms to Solids, from Lasers to Intense X-rays}}},\ \bibinfo {editor}
  {edited by\ \bibinfo {editor} {\bibfnamefont {M.}~\bibnamefont {Kitzler}}\
  and\ \bibinfo {editor} {\bibfnamefont {S.}~\bibnamefont {Gr{\"a}fe}}}\
  (\bibinfo  {publisher} {Springer International Publishing},\ \bibinfo
  {address} {Cham},\ \bibinfo {year} {2016})\ pp.\ \bibinfo {pages}
  {3--25}\BibitemShut {NoStop}%
\bibitem [{\citenamefont {Dubois}\ \emph {et~al.}(2018)\citenamefont {Dubois},
  \citenamefont {Berman}, \citenamefont {Chandre},\ and\ \citenamefont
  {Uzer}}]{Dubois2018}%
  \BibitemOpen
  \bibfield  {author} {\bibinfo {author} {\bibfnamefont {J.}~\bibnamefont
  {Dubois}}, \bibinfo {author} {\bibfnamefont {S.~A.}\ \bibnamefont {Berman}},
  \bibinfo {author} {\bibfnamefont {C.}~\bibnamefont {Chandre}},\ and\ \bibinfo
  {author} {\bibfnamefont {T.}~\bibnamefont {Uzer}},\ }\bibfield  {title}
  {\bibinfo {title} {Capturing photoelectron motion with guiding centers},\
  }\href {https://doi.org/10.1103/PhysRevLett.121.113202} {\bibfield  {journal}
  {\bibinfo  {journal} {Phys. Rev. Lett.}\ }\textbf {\bibinfo {volume} {121}},\
  \bibinfo {pages} {113202} (\bibinfo {year} {2018})}\BibitemShut {NoStop}%
\bibitem [{\citenamefont {Zhao}\ \emph {et~al.}(2019)\citenamefont {Zhao},
  \citenamefont {Zhou}, \citenamefont {Liang}, \citenamefont {Zeng},
  \citenamefont {Ke}, \citenamefont {Liu}, \citenamefont {Li},\ and\
  \citenamefont {Lu}}]{Zhao2019}%
  \BibitemOpen
  \bibfield  {author} {\bibinfo {author} {\bibfnamefont {Y.}~\bibnamefont
  {Zhao}}, \bibinfo {author} {\bibfnamefont {Y.}~\bibnamefont {Zhou}}, \bibinfo
  {author} {\bibfnamefont {J.}~\bibnamefont {Liang}}, \bibinfo {author}
  {\bibfnamefont {Z.}~\bibnamefont {Zeng}}, \bibinfo {author} {\bibfnamefont
  {Q.}~\bibnamefont {Ke}}, \bibinfo {author} {\bibfnamefont {Y.}~\bibnamefont
  {Liu}}, \bibinfo {author} {\bibfnamefont {M.}~\bibnamefont {Li}},\ and\
  \bibinfo {author} {\bibfnamefont {P.}~\bibnamefont {Lu}},\ }\bibfield
  {title} {\bibinfo {title} {Frustrated tunneling ionization in the
  elliptically polarized strong laser fields},\ }\href
  {https://doi.org/10.1364/OE.27.021689} {\bibfield  {journal} {\bibinfo
  {journal} {Opt. Express}\ }\textbf {\bibinfo {volume} {27}},\ \bibinfo
  {pages} {21689} (\bibinfo {year} {2019})}\BibitemShut {NoStop}%
\end{thebibliography}
\end{document}